\documentclass[pra,twocolumn,amsmath,amssymb,groupaddress,eqsecnum]{revtex4-2}
\usepackage{graphicx,amsmath,relsize,epstopdf,color,mathtools,bm,newtxtext,newtxmath,braket,rotating,float}
\usepackage[hyphenbreaks]{breakurl}
\usepackage[colorlinks=true,linkcolor=blue,citecolor=blue,urlcolor =blue]{hyperref}
\usepackage[normalem]{ulem}
\usepackage[table,xcdraw]{xcolor}
\usepackage{braket}

\newcommand{\Tr}{\mathop{\mathrm{Tr}} \nolimits}

\newcommand{\matriz}[1]{\mathsf{#1}}

\begin{document}
\title{Neural-network quantum state tomography}

\author{Dominik~Koutn\'{y}}
\affiliation{Department of Optics, Palacky University, 17. listopadu 12, 77146 Olomouc, Czech Republic}
\author{Libor~Motka}
\affiliation{Department of Optics, Palacky University, 17. listopadu 12, 77146 Olomouc, Czech Republic}

\author{Zden\v{e}k~Hradil}
\affiliation{Department of Optics, Palacky University, 17. listopadu 12, 77146 Olomouc, Czech Republic}

\author{Jaroslav~\v{R}eh\'{a}\v{c}ek} 
\affiliation{Department of Optics, Palacky University, 17. listopadu 12, 77146 Olomouc, Czech Republic}

\author{L.~L.~S\'{a}nchez-Soto}
\affiliation{Departamento de \'Optica, Facultad de F\'{\i}sica, Universidad Complutense, 28040~Madrid, Spain}
\affiliation{Max-Planck-Institut f\"ur die Physik des Lichts,  91058~Erlangen, Germany}

\begin{abstract}
We revisit the application of neural networks techniques to quantum state tomography.  We confirm that the positivity constraint can be successfully implemented with trained networks that convert outputs from standard feed-forward neural networks to valid descriptions of quantum states. Any standard neural-network architecture can be adapted with our method.  Our results  open possibilities to use state-of-the-art deep-learning methods for quantum state reconstruction under various types of noise. 
\end{abstract}

\maketitle

\section{Introduction}

Modern quantum technologies exploit distinctive features of quantum systems   to achieve performances unattainable by classical strategies. This potential advantage hinges on the capability to create, manipulate, and measure quantum states. Any experimental procedure in this area requires a reliable certification of these steps: this is precisely the province of quantum state tomography (QST)~\cite{lnp:2004uq}.

The goal of QST is to estimate the unknown quantum state through measurements performed on a finite set of identical copies of the system. If the state is described by the density matrix $\varrho$, living in a $d$-dimensional Hilbert space,  $O(d/\varepsilon)$ copies are required to obtain an estimate of $\varrho$  with an error (understood as total variation distance) less than $\varepsilon$~\cite{ODonnell:2016wk}. This clearly illustrates the resource requirements of QST for large-scale systems. 

In a broad sense, QST is an inverse problem~\cite{Artiles:2005wk,Tarantola:2005aa,Aster:2013aa}. As such, the linear inversion~\cite{Qi:2013ts} is probably the most intuitive approach to the topic. Yet it has some cons too: it might report a nonphysical state and the mean squared error bound of the estimate cannot be determined analytically. To bypass these drawbacks a variety of useful QST methods, such as Bayesian tomography~\cite{Blume-Kohout:2010uf,Granade:2016tr}, compressed sensing~\cite{Gross:2010aa,Ahn:2019uw}, or matrix-product states~\cite{Cramer:2010aa,Lanyon:2017vi}, are at hand, although the  maximum-likelihood estimation (MLE) is still the most common approach~\cite{Hradil:1997tt,James:2001ut}.

From a modern perspective, QST is fundamentally a data processing problem, trying to extract information from as few noisy measurements as possible. Therefore, the estimation algorithms used in QST can be easily translated into tasks in machine learning (ML)~\cite{Murphy:2012uq,Nielsen:2018ul,Mehta:2019tt}.  Actually, neural networks (NNs) have been used to address data-driven problems: examples in quantum information include identifying phase transitions~\cite{Rem:2019vc}, detecting nonclassical features~\cite{Gebhart:2020tx,You:2020uj,Harney:2020ta}, quantum error correction~\cite{Torlai:2017tz,Baireuther:2018tj,Krastanov:2017vr,Fitzek:2020ux}, calibrating quantum devices~\cite{Flurin:2020vf,Wittler:2021vj}, speeding up quantum optimal control~\cite{Leung:2017tx}, and designing quantum experiments~\cite{Krenn:2016te,Melnikov:2018tv,ODriscoll:2019wz}, to cite only but a few. 

Recently, ML has been applied to QST with very promising results~\cite{Torlai:2018wn,Carleo:2018tj,Xin:2019vo,Palmieri:2020vc,Melkani:2020wo,Lohani:2020vi,Liu:2020tn,Quek:2021uc,Carrasquilla:2021tx}. In particular, generative models~\cite{Carrasquilla:2019wm,Lloyd:2018uw}, usually restricted Boltzmann machines, have been used to treat the measurement outcomes on a quantum state~\cite{Tiunov:2020wm}. These are NNs containing two layers, visible and hidden, with all-to-all connections between the neurons in different layers and none inside each layer. This technique, although powerful, suffers from difficulties with sampling and a lack of straightforward training for larger models.

The use of feed-forward architectures, including recurrent NNs, has been recently advocated~\cite{Cai:2018wn,Cha:2021wn} because these architectures are easier to train without any need for sampling steps, using gradient-based optimization with backpropagation. However, generative tasks in ML often use variational autoencoders~\cite{Kingma:2019uw} and generative adversarial NNs~\cite{Goodfellow:2014td}. These are now being actively explored for learning quantum states~\cite{Rocchetto:2018wg,Zoufal:2019wi,Ahmed:2021vs,Ahmed:2021vy}.

Our motivation in this paper is to address the benefits of NN-based reconstruction over standard techniques.  To fairly benchmark the performance we pick three representative estimators, namely linear inversion, its positivized version, and MLE, and compare them with a typical NN estimator, obtained with a feed-forward architecture. As measurement, we choose the so-called square root measurement, which was introduced as a ``pretty good measurement"~\cite{Hausladen:1994aa} for distinguishing possibly nonorthogonal states. Using the Hilbert-Schmidt distance between the true and the reported states as our main indicator, our results suggest that NNs predict unknown quantum states about three orders of magnitude faster compared to linear and MLE estimates. Interestingly, the average errors are similar for all the estimators considered in all dimensions. This confirms the power of deep-learning-based tools for the quantum realm.

This paper is organized as follows. In Sec.~\ref{sec:essentials}, we briefly discuss the basic tools of QST we need for our purposes.   In Sec.~\ref{sec:NNs} we describe the details of our NN architecture and training methods. Then, we present the performance of the different estimators in Sec.~\ref{sec:results}, while our conclusions are summarized in Sec.~\ref{sec:conc}.

\section{Background}

\label{sec:essentials}

We first set the stage for our model. We shall be considering a $d$-dimensional quantum system, described by a $d\times d$ density matrix $\varrho$, which requires $n \equiv d^{2} - 1$ independent real numbers for its specification.  

The goal of QST is to estimate $\varrho$ from measurements performed on identically prepared copies of the system. These measurements are, in general, represented by positive operator-valued measures (POVMs)~\cite{Helstrom:1976ij}: they are a set of positive Hermitian operators $\{ \Pi_{\ell} \}$, with the properties
\begin{equation}
\Pi_{\ell} \geq 0 \, ,  \qquad 
\Pi_{\ell}^{\dagger} = \Pi_{\ell} \, , \qquad
\sum_{\ell} \Pi_{\ell} = \openone \, .
\end{equation} 
Each POVM element represents a single output of the measuring apparatus. We take every measurement as yielding $m$ distinct outcomes (which we assume to be discrete). According to Born's rule, the probability of detecting the $\ell$th output is given by 
\begin{equation}
p_{\ell} = \Tr (\varrho \Pi_{\ell}) \, .   
\end{equation}

To invert this equation, it is convenient to map both $\varrho$ and  $\{ \Pi_{\ell} \}$ into a suitable vector form. To this end, we use a traceless Hermitian operator basis $\{\Gamma_{k}\}$ ($k=0, \ldots, n$)  and $\Gamma_0= \openone$,  satisfying $\Tr (\Gamma_{j})= 0$ and $\Tr (\Gamma_{j} \Gamma_{k}) = \delta_{jk}$. In this way, we get the parametrization
\begin{equation}
  \label{rhodecomp}
  \varrho = r_{0} \, \Gamma_{0} + \sum_{k=1}^{n} r_{k} \, \Gamma_{k} \, \qquad \qquad
   \Pi_{\ell} = w_{\ell} + \sum_{k=1}^{n} \matriz{C}_{\ell k} \;  \Gamma_{k} \, .
\end{equation}
Although the condition $\Tr \varrho = 1$ sets $r_{0} = 1/d$, we leave $r_{0}$ as a parameter to keep the same number of unknowns as for the approach using Cholesky decomposition to be described later. The important point is that the state is characterized by the Bloch vector~\cite{Kimura:2003wz,Bertlmann:2008uh,Mendas:2008uf,Bruning:2012vd} $r_{k} = \Tr (\varrho \Gamma_{k})$, whereas $\matriz{C}_{\ell k} = \Tr ( \Pi_{\ell} \Gamma_{k})$ is a $ m \times n$ real matrix describing  the explicit relation between the theoretical probabilities $\mathbf{p}$ and the state parameters $\mathbf{r}$.  

In consequence, the inverse problem we have to solve turns out to be the linear system
\begin{equation} 
\label{eq:LinTomo}
\mathbf{p} = \matriz{C} \, \mathbf{r} \, ,
\end{equation}
where we have omitted an unessential constant term that can be incorporated into the following discussion in a straightforward way.  

In presence of noise and with a finite number of copies the collected data, we will denote by $\mathbf{f}$, deviates from the expected values $\mathbf{p}$. The ultimate goal of QST is to infer the   signal parameters $\mathbf{r}$ from the measured noisy data $\mathbf{f}$.   A naive solution is to use the estimator
\begin{equation}
\widehat{\mathbf{r}}_{\textsc{LI}} = \matriz{C}^{-} \; \mathbf{f} \, ,
\end{equation} 
where the $\matriz{C}^{-}$ stands for pseudoinverse~\cite{Penrose:1955aa,Ben-Israel:1977aa,Campbell:1991aa} and the subscript LI reminds us that this is a linear inversion approach.  

This $\widehat{\mathbf{r}}_{\mathrm{LI}}$ is also known as the ordinary least squares estimator~\cite{Lawson:1974aa}. As heralded before, the resulting $\widehat{\mathbf{r}}_{\mathrm{LI}}$ is no longer guaranteed to represent a positive semidefinite operator.  One might ensure positivity by setting the negative eigenvalues to zero, which has been called the ``quick and dirty" approach~\cite{Kaznady:2009wf}, although this performs poorly.

Another alternative is to use instead the generalized least-square estimator~\cite{Opatrny:1997vv}, defined as $\widehat{r}_{\mathrm{GLS}} = (\matriz{X}^{-1} \matriz{C})^{+} \matriz{X}^{-1} ) \mathbf{f}$, where $\matriz{X}$ is such that $\matriz{C} \matriz{C}^{\dagger}$ is the data covariance matrix.  Under the Gauss-Markov assumptions~\cite{Hallin:2006aa} it is  the best linear unbiased estimator (usually  known as \textsc{blue})~\cite{Kay:1993aa}.  However, for small and medium sized data sets, a reliable estimation of the data covariances is not possible, and then $\widehat{\mathbf{r}}_{\mathrm{LI}}$ turns out to be a handy estimator.

To circumvent these obstructions we might follow yet another route, introducing instead a semidefinite program that solves \eqref{eq:LinTomo}, together with the positivity constraint. The resulting estimator, denoted by $\widehat{\mathbf{r}}_{\mathrm{SDP}}$, is thus a solution of
\begin{equation}
\label{eq:SDPTomo}
\begin{array}{ll}
\mathrm{minimize} &   ||\mathbf{f} - \matriz{C} \mathbf{r}|| \\ 
& \\
 \mathrm{subject \ to} & \varrho \geq 0 \quad \mathrm{and} \quad \Tr \varrho =1.
 \end{array}
\end{equation}

Finally, to make our analysis complete and consider a nonlinear estimator, we also incorporate the MLE, which guarantees positivity of the resulting quantum state. Although there is a vast literature on the subject, the MLE estimate $\widehat{\varrho}_{\mathrm{MLE}}$ can be seen as the fixed point of the iterative map~\cite{Rehacek:2007wr}
\begin{equation}
\label{eq:MaxLikTomo}
\varrho_{k+1} \leftarrow {\lambda_k} R \varrho_{k} R,
\end{equation}
with 
\begin{equation}
R = \sum_j \frac{f_j}{p_j} \Pi_j
\end{equation} and $\lambda_k$ is a normalization constant. The resulting Bloch-vector $\widehat{\mathbf{r}}_{\mathrm{MLE}}$ estimate is asymptotically unbiased as $f_j \rightarrow p_j$. Usually, a few thousands of iterations are needed to observe the stationary point of the map \eqref{eq:MaxLikTomo}.

From a numerical point of view, an efficient way to deal with the positivity constraint is to directly decompose the density operator using the famous Cholesky factorization~\cite{Watkins:1991tc} 
\begin{equation}
\varrho = \frac{\matriz{A} \matriz{A}^{\dagger}}{\Tr ( \matriz{A} \matriz{A}^{\dagger})} 
 \, ,
\end{equation} 
where $\matriz{A}$ is a complex lower triangular matrix and $\matriz{A}^{\dagger}$ its Hermitian conjugate.  Born's rule then turns to a set of nonlinear equations, which are rather complicated to solve. For this purpose, we adopt ML techniques.

\section{Neural-network estimators}
\label{sec:NNs}

Our goal is to built a NN that links the input observed frequencies $\mathbf{f}$ to either an output true Bloch vector $\widehat{\mathbf{r}}_{\mathrm{NN}}$ or a Cholesky matrix $\matriz{A}_{\mathrm{NN}}$. The sampled frequencies $\mathbf{f}$ serve as input to the NN, which transforms them into an output Bloch vector or a Cholesky matrix via a series of linear transformations, each followed by evaluation of some nonlinear function. The structure of such transformations is represented by neurons ordered into deep layers. More precisely, the values $\mathbf{z}^{(k)}$ of the neurons in $k$th layer are
\begin{equation}
\label{eq:DNNlayer}
\begin{array}{ll}
\mathbf{z}^{(k)} & = {f \left (\mathbf{y}^{(k)} \right )},   \\
& \\
\mathbf{y}^{(k)} & = \matriz{W}^{(k-1\rightarrow k)}\mathbf{z}^{(k-1)}+\mathbf{b}^{(k-1)}, 
\end{array}
\end{equation}
where $\matriz{W}$ is a matrix of weights connecting neighboring layers that together with the vector of biases $\mathbf{b}$ forms a set of trainable parameters. The nonlinear activation scalar function $f$ can be chosen arbitrarily, depending on the problem. In our case, the rectified linear unit function $f_{\mathrm{ReLU}} (x) :=  \max(0,x)$  is used in every deep layer except at the output layer, where we take $f(x) = \tanh(x)$. The hyperbolic tangent function maps real numbers into the $(-1,1)$ interval which coincides with restriction on elements of both Bloch vector and Cholesky matrix. Symbolically, we can express the Bloch vector (and similarly  the elements of the Cholesky matrix) as
\begin{equation}
\label{eq:StokesDNN}
	\mathbf{r}_{\mathrm{NN}} = f_{\tanh} \circ \widetilde{\matriz{W}}^{(n-1 \rightarrow \mathrm{out})}\circ \cdots \circ f_{\mathrm{ReLU}} \circ \widetilde{\matriz{W}}^{(\mathrm{in}\rightarrow 1)}\circ \mathbf{f},
\end{equation}
where $\widetilde{\matriz{W}}$ is a shortcut for $\{\matriz{W},\mathbf{b}\}$. 

\begin{figure}[t]
  \centering
  \includegraphics[width=0.90\columnwidth]{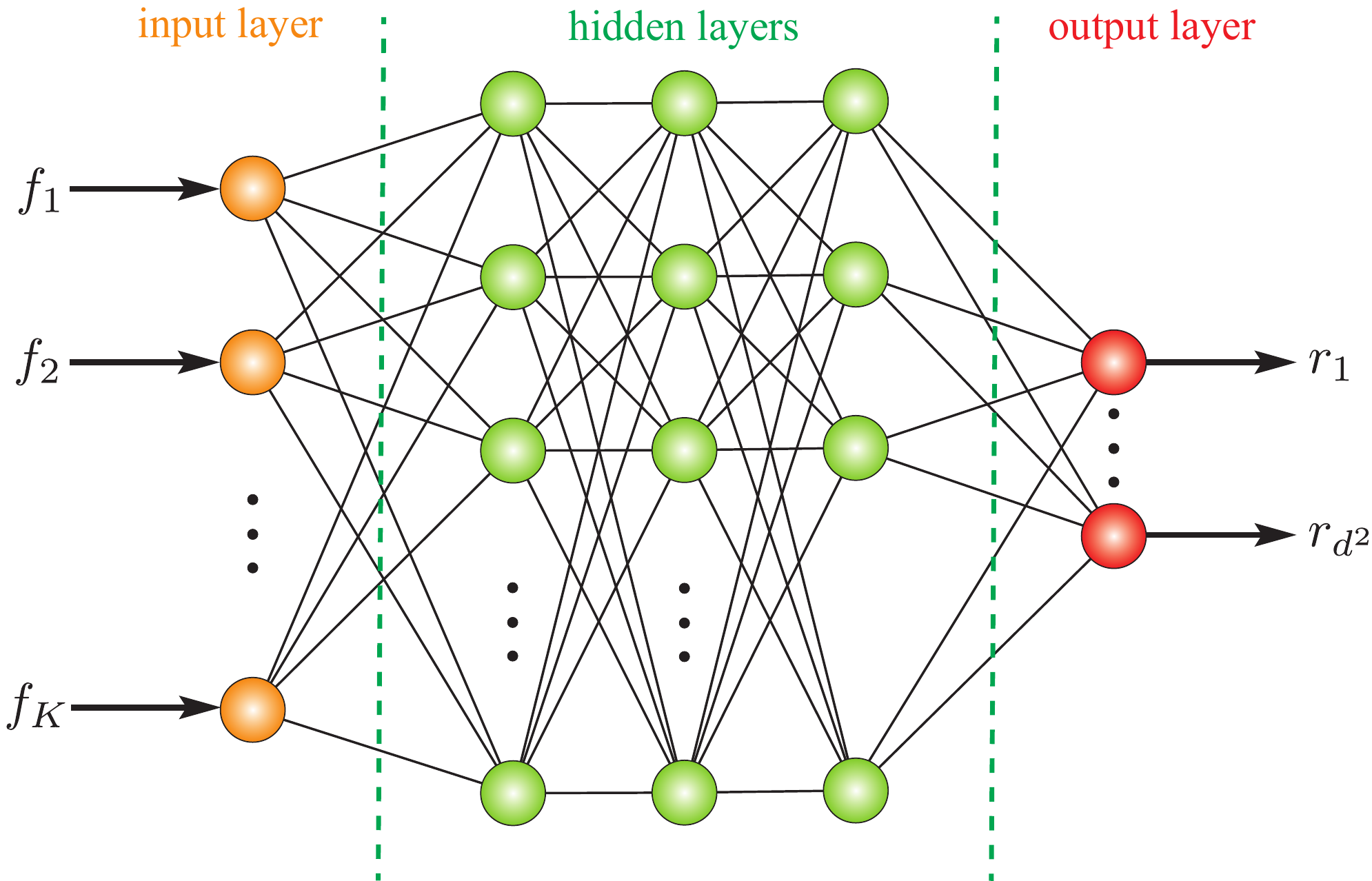}
  \caption{Sketch of a NN for estimating the Bloch  parameters or the elements of the Cholesky matrix from the sampled statistics $\mathbf{f}$. The filled blue ovals represent neurons in deep layers, representing about $2\times 10^5$ trainable parameters. The output layer has a hyperbolic tangent as an activation function since it has the same bounds as the Bloch parameters and theelements of the Cholesky matrix. In the hidden layers, the rectified linear unit is used as an activation function. The structure of the NN is  the same,  independently of the dimension or specific parametric representation of quantum states. }
  \label{fig:netSketch}
\end{figure}

The NN learns by minimizing the loss function. We chose to work with mean squared error, which takes the form
\begin{equation}
\label{eq:loss}
\mathcal{L}= \left \langle \sum_{k=0}^{n} \left | r_{k}-z_{k}^{\mathrm{out}} ( \widetilde{\matriz{W}},f )  \right |^2\right\rangle \, ,
\end{equation}
where $\langle \cdot \rangle$ denotes the average value in the state $\varrho$. Optimization in the NN is done by backpropagating the error. This is arguably the workhorse of most ML algorithms and definitely the standard approach in most situations, which is working with batches of data.  The term and its general use in NNs was coined in~\cite{Rumelhart:1986we} and a modern overview is given in textbook~\cite{Goodfellow:2016wc}.

We minimize the value of the loss function $\mathcal{L}$ over all components of a given dataset to update weights and biases $\widetilde{\matriz{W}}$, using a stochastic gradient-based optimization, which is of core practical importance in many fields~\cite{Ruder:2016tr}. A widely accepted algorithm is Adam~\cite{Kingma:2014us}, which is straightforward to implement,  computationally efficient, and has little memory requirements. We use an improved version that incorporates Nesterov-accelerated adaptive momentum estimation (Nadam)~\cite{Dozat:2015wi}, since recent results indicate that it has better performance~\cite{Dogo:2018vp}.  At the step $t$, the Nadam procedure updates parameters in the form
\begin{eqnarray}
{\widetilde{\mathbf{W}}_t \leftarrow \widetilde{\mathbf{W}}_{t-1} - \eta \frac{\bar{\mathbf{m}}_t}{\sqrt{\hat{\mathbf{n}}_t}+\epsilon} \, ,} 
\end{eqnarray}
with
\begin{eqnarray}
\mathbf{g}_t & \leftarrow & \nabla_{ \widetilde{\mathbf{W}}_{t-1}} \mathcal{L}\left(\widetilde{\mathbf{W}}_{t-1}\right), \nonumber \\
\mathbf{\hat{g}} & \leftarrow & \frac{\mathbf{g}_t}{1-\prod_{j=1}^{t}\mu_j}, \nonumber \\
\mathbf{m}_t & \leftarrow & \mu \mathbf{m}_{t-1}+(1-\mu)\mathbf{g}_t,  \nonumber \\
\mathbf{\hat{m}}_t & \leftarrow & \frac{\mathbf{m}_t}{1-\prod_{j=1}^{t+1}\mu_j},  \\
\mathbf{n}_t & \leftarrow & \nu \mathbf{n}_{t-1}+(1-\nu)\mathbf{g}_t^2, \nonumber\\
\mathbf{\hat{n}}_t & \leftarrow & \frac{\mathbf{n}_t}{1-\nu^t}, \nonumber \\
\mathbf{\bar{m}}_t & \leftarrow & (1-\mu_t) \mathbf{\hat{g}}_t +\mu_{t+1}\mathbf{\hat{m}}_t. \nonumber 
\end{eqnarray}
Here, $\eta$ represents the learning rate, $\mu$ the exponential decay rate for the first moment estimates $\mathbf{\hat{m}}$,  $\nu$ the exponential decay rate for the weighted norm $\mathbf{g}_t^2$, and $\epsilon$ is a parameter that ensures the numerical stability of the Nadam optimization procedure. We set the numerical values   $\{\eta,\mu,\nu,\epsilon\}$ to $\{0.001,0.9,0.999,10^{-7}\}$.

All the above is implemented in Keras~\cite{Chollet:2015wy} and Tensorflow~\cite{Abadi:2015te} libraries for Python. The corresponding code can be found in Ref.~\cite{Koutny:2022qp}. In every epoch, the training dataset is divided into $100$ batches. The number of epochs needed to find a global minimum of the loss function varies across different deep NNs. In general, the training stopped after  $400-2000$ epochs, depending on the dimension in which we estimate quantum states.  We defined an early stopping  after not finding the better minimum of the loss function in the $200$ consecutive epochs. We stress that both $\widehat{\mathbf{r}}_{\mathrm{NN}}$ and $\widehat{\mathbf{r}}_{\mathrm{LI}}$ estimates and the ensuing quantum states are Hermitian matrices but do not incorporate the positivity constraint, in contradiction with  $\widehat{\mathbf{r}}_{\mathrm{SDP}}$ and $\widehat{\mathbf{r}}_{\mathrm{MLE}}$.

\begin{figure*}[t]
  \centering
  \includegraphics[width=1.85\columnwidth]{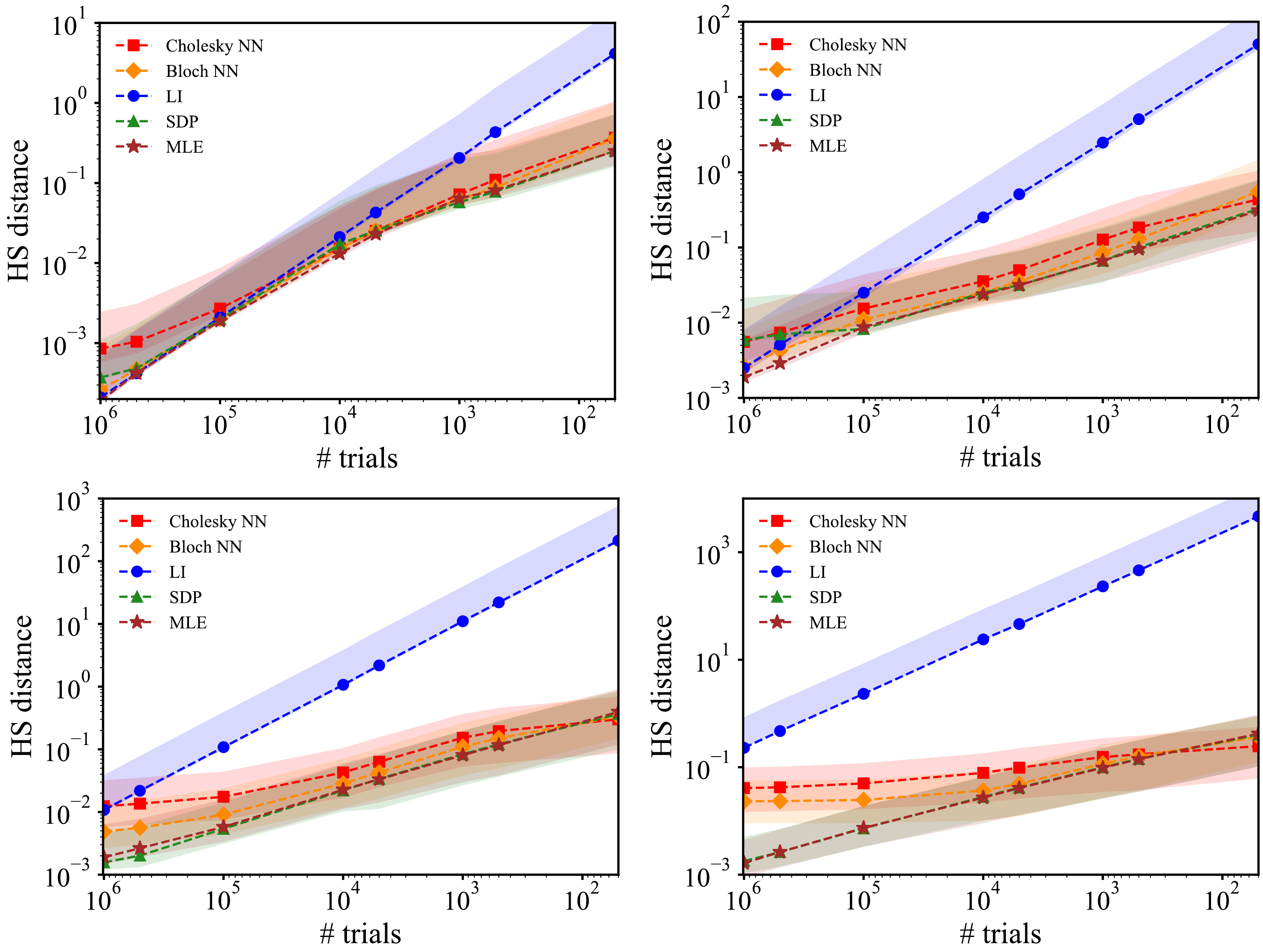}
  \caption{Average Hilbert-Schmidt distance for different estimating strategies on the number of trials with which we sampled the true probability distribution $\mathbf{p}$ in dimension 3 (upper left panel), 5 (upper rigt panel), 7 (lower left panel), and 9 (upper right panel). The insets indicate the corresponding estimators. The average errors for both NN in the undersampled regime are of the same order for SDP and MLE in all dimensions. When using highly sampled statistics, number of trials $>10^5$, MLE starts to outperfonm both NN approaches. Confidence intervals of 80\% are depicted in respective colors.}
    \label{fig:HSerrors}
\end{figure*}

\begin{figure}[t]
  \centering
  \includegraphics[width=\columnwidth]{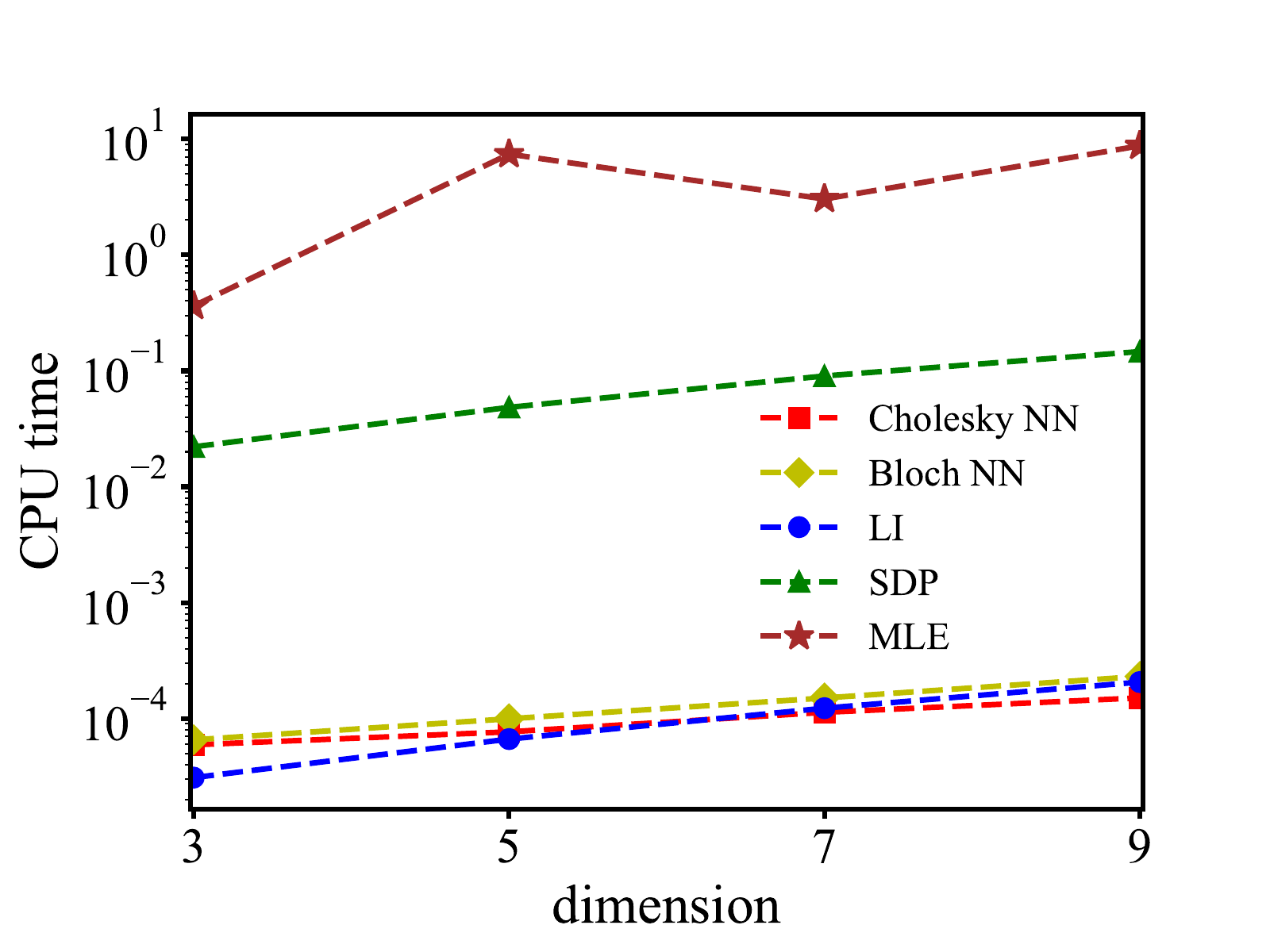}
  \caption{Average time per state estimation for different tomographical methods, as indicated in the inset. The NN approach is about three orders faster than SDP  and about four order faster than  MLE.}
  \label{fig:Time}
\end{figure}

\section{Results}
\label{sec:results}

Our deep NN is built as follows:  The input layer is fed by observed frequencies $\mathbf{f}$ for different quantum states, followed by eight layers consisting of $(200,180^{\otimes 2},160^{\otimes 4},100)$ neurons with the ReLU activation function. The output layer, with a hyperbolic tangent activation function, serves as an estimate of Bloch vector or the elements of Cholesky matrix. 

The structure of the NN was adjusted heuristically, after having tried multiple settings with differing number of free parameters and deep layers. In terms of the distance between estimated and true quantum states, we got on average the same accuracy for a NN with two layers. However,   it was observed that NNs are more likely to return parameters corresponding to positive semidefinite matrices, compared to, e.g., the LI method.

The NN sketch is presented in Fig.~\ref{fig:netSketch}. We trained in total eight NNs, each with the same structure, for the inference of quantum states in dimensions $d = 3,5,7,$ and $9$. As our target states, we use random density matrices $\varrho$  distributed according to~\cite{Osipov:2010ux}
\begin{equation}
\label{eq:randomState}
{\varrho = \frac{\matriz{X}\matriz{X}^{\dagger}}{\Tr (\matriz{X}\matriz{X}^{\dagger})} , }
\end{equation}
with $\matriz{X}$ pertaining to the Ginibre ensemble~\cite{Ginibre:1965vc}, that is, with real and imaginary parts of each matrix entry being independent normal random variables. These are implemented in Python using \textsc{QuTiP}~\cite{Johansson:2012wf}. 

As heralded in the Introduction, as our measurement scheme, we choose the square root measurement, defined by the rank-one POVM
\begin{equation}
\label{eq:srm}
{\Pi_{\ell} = G^{-1/2} \; \ket{\phi_{\ell}} \bra{\phi_{\ell}} \, G^{-1/2} \, , 
\qquad 
G = \sum_\ell \ket{\phi_\ell}\bra{\phi_\ell} \, ,}
\end{equation}
where $|\phi_{\ell} \rangle$ are randomly generated Haar-distributed pure states~\cite{Zyczkowski:2011wo} $(\ell=0, \ldots, n)$.  This POVM is known to be optimal, in the sense that the measurement vectors are the closest in the squared norm to the given states~\cite{Eldar:2001uw,Dalla-Pozza:2015aa}.

\begin{figure*}[t]
  \centering
  \includegraphics[width=1.80\columnwidth]{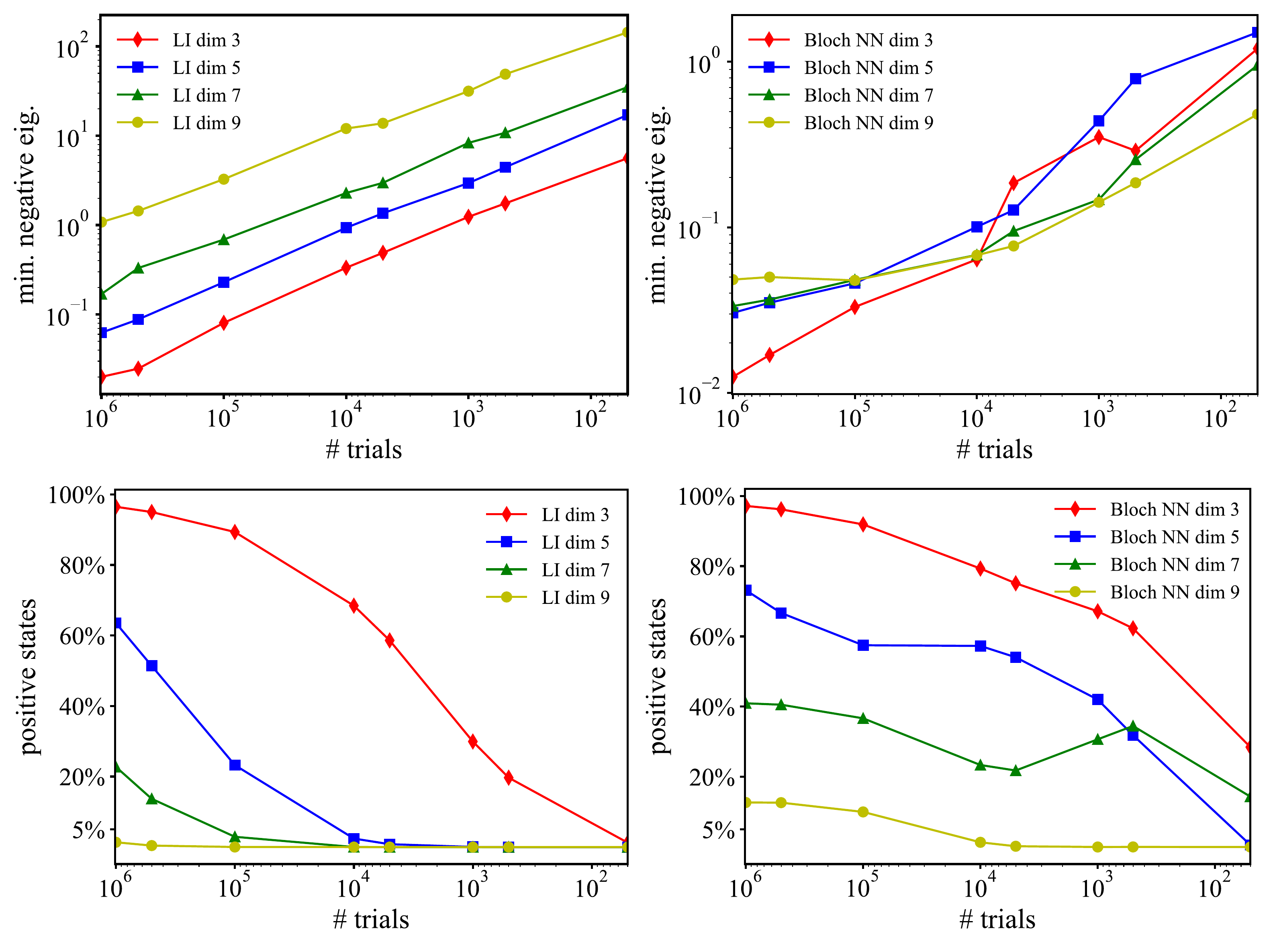}
  \caption{Quality of estimators based on linear inversion and NN. In all the 
  panels, red diamonds, blue squares, green triangles, and yellow circles represent the results for estimators in dimensions 3, 5, 7, and 9, respectively.  The upper plots show the dependence of the most negative eigenvalue on the number of trials for those dimensions and for the LI estimator (left) and the Bloch NN estimator (right). The bottom panels depict the percentage of positive semidefinite states amongst all the reconstructed states with LI (left) and Bloch NN (right). The results are obtained from datasets containing $10^{4}$ statistics in each dimension.}
  \label{fig:LeastNegEig}
\end{figure*}

For each dimension, the training (validation) dataset contains $8\times 10^{5}\, (2\times 10^{5})$  points.  One quarter of the data in the training dataset are probabilities $p_{\ell} = \Tr (\varrho \Pi_{\ell} )$ sampled with a random number of trials, ranging from $d^2$ up to $10^5$, the rest are theoretical probabilities. Each input vector, containing either theoretical or sampled statistics, corresponds to different randomly generated quantum states \eqref{eq:randomState}, whereas the measurement \eqref{eq:srm} is fixed for all states in each dimension.

Ideally, all data points in the training dataset should consist of only theoretical probabilities so that the NN can extract the appropriate transformation. However, in the undersampled regime, it turns out to be beneficial for the NN to see examples of the sampled statistics. In this case, with a training dataset containing only theoretical probabilities, the NN would lose the ability to predict positive matrices, while with only sampled probabilities in the training dataset, the NN would have a hard time in correctly learning the mapping from probabilities to the Bloch vector or the Cholesky matrix. 

The output layer consists of true values of the Bloch vector or elements of the Cholesky matrix. All NNs are trained for at most 2000 epochs, which takes about 12 hours for every NN when estimating density matrices in dimension 9. After the training procedure, we compared estimates of quantum states based on NNs with  standard methods, namely LI, SPD, and MLE.

Results are shown in Fig.~\ref{fig:HSerrors}. For each dimension $d$, we generated $10^{3}$ random density matrices and, using the same measurement scheme as for the training, we obtained a set of $10^{3}$ probability distributions, often called  a test set. After getting the test set, we sampled each probability distribution with a number of trials depicted on the horizontal axis. Then, we used trained NNs, LI, MLE and the SDP to reconstruct density matrices from the sampled statistics. We plot the average Hilbert-Schmidt distance between true and inferred quantum states as a function of the number of trials. As we can see, NNs outperform the $\widehat{\mathbf{r}}_{\mathrm{LI}}$ estimator and are better or give errors in the same order of magnitude as $\widehat{\mathbf{r}}_{\mathrm{SPD}}$. 

For a fixed number of trials, the NN-based Bloch vector has a better average error compared to the Cholesky one,  but tends to report Hermitian matrices with nonpositive least eigenvalues. Moreover, NNs work relatively  well in the undersampled regime. Interestingly, NNs show the ability to extrapolate beyond the number of samples on which they were trained. This can be appreciated in the lower Hilbert-Schmidt distance  for $10^6$ trials, when only up to $10^5$ trials were used to sample the true statistics. 

The combination of theoretical and sampled probabilities in the training set was balanced in such a way that the NNs work well in the undersampled regime, but also provide accurate estimates when the number of trials is high. Of course, when the number of trials goes to infinity; i.e, when working with theoretical probabilities, the $\widehat{\mathbf{r}}_{\mathrm{LI}}$ and $\widehat{\mathbf{r}}_{\mathrm{MLE}}$ estimators converge to the true state of the system.

Figure~\ref{fig:Time} shows an analysis of the performance of different estimators. We depict the average time per single evaluation of $\widehat{\mathbf{r}}_{\mathrm{LI}}$, $\widehat{\mathbf{r}}_{\mathrm{SDP}}$, $\widehat{\mathbf{r}}_{\mathrm{MLE}}$, and both NNs. For the  NNs, the times shown are only those associated with the prediction phase, not the training (which takes a lot longer). The semidefinite program infers the quantum state from sampling statistics at around $10^{-1}$ seconds. MLE turns out to be the most time-consuming procedure. Compared to linear inversion and both NN approaches, MLE predicts quantum states about $10^4$ slower. Akin to linear inversion, NN predicts unknown quantum state from the data about three orders faster compared to SPD estimates and about four orders faster compared to the MLE estimates.

Figure~\ref{fig:LeastNegEig} summarizes our performance analysis. We compare the quality of $\widehat{\mathbf{r}}_{\mathrm{LI}}$ and $\widehat{\mathbf{r}}_{\mathrm{NN}}$ in terms of the \emph{quantumness} of the inferred states. We show  the mean of the largest negative eigenvalues and ratio of positive semidefinite quantum states among the set of estimated Hermitian matrices on the measured statistics sampled with given number of trials. Note that we have excluded results from  MLE and SDP, for those estimators always reconstruct a positive matrix.

As one can see in Fig.~\ref{fig:LeastNegEig}, NNs  can learn the positivity constraint. For example, in dimension 9, considering $\widehat{\mathbf{r}}_{\mathrm{LI}}$, only $1\%$ of all Hermitian matrices are positive semidefinite, compared to $17\%$ using NN, estimated from statistics sampled with $10^6$ trials. In the undersampled regime, where the number of trials is in the order of the number of projectors, NNs in each dimension predict the higher number of positive quantum states compared to the $\widehat{\mathbf{r}}_{\mathrm{LI}}$  estimator.

The issue with predicting quantum states with negative eigenvalues is that it is more prevalent for states that are singular (i.e., with vanishing determinant), as these are the states that sit on the boundary in the generalized Bloch representation. As such, the results in Fig.~\ref{fig:LeastNegEig} depend 
on the purity  of the states to be reported.

\section{Concluding remarks}
\label{sec:conc}

In summary, we have shown how NN can assist in the reconstruction of quantum states.  The NN maps the input experimental data to a valid density matrix up to three orders of magnitude faster than the standard QST. This presents a significant advantage for data postprocessing during tomography. The NN learns to represent the state in a way that is well suited for the problem. 

Our results confirm how some of the latest ideas from deep learning can be quite easily adapted and applied to quantum information tasks with just a few tweaks to incorporate the rules of quantum physics. This opens up a wealth of possible applications, which are the object of intense investigation.

\section*{Ackowledgments}

The authors thank Miroslav Je\v{z}ek for useful discussions and two anonymous reviewers for their constructive and detailed comments. This work was supported by the European Union's Horizon 2020 research and innovation programme under the QuantERA programme through the project ApresSF and from the EU Grant  899587 (Project Stormytune), the Palack\'y University Grant {IGA}\_PrF\_2021\_002 and the Spanish Ministerio de Ciencia e Innovacion (Grant PGC2018-099183-B-I00).


\begin{thebibliography}{82}%
\makeatletter
\providecommand \@ifxundefined [1]{%
 \@ifx{#1\undefined}
}%
\providecommand \@ifnum [1]{%
 \ifnum #1\expandafter \@firstoftwo
 \else \expandafter \@secondoftwo
 \fi
}%
\providecommand \@ifx [1]{%
 \ifx #1\expandafter \@firstoftwo
 \else \expandafter \@secondoftwo
 \fi
}%
\providecommand \natexlab [1]{#1}%
\providecommand \enquote  [1]{``#1''}%
\providecommand \bibnamefont  [1]{#1}%
\providecommand \bibfnamefont [1]{#1}%
\providecommand \citenamefont [1]{#1}%
\providecommand \href@noop [0]{\@secondoftwo}%
\providecommand \href [0]{\begingroup \@sanitize@url \@href}%
\providecommand \@href[1]{\@@startlink{#1}\@@href}%
\providecommand \@@href[1]{\endgroup#1\@@endlink}%
\providecommand \@sanitize@url [0]{\catcode `\\12\catcode `\$12\catcode
  `\&12\catcode `\#12\catcode `\^12\catcode `\_12\catcode `\%12\relax}%
\providecommand \@@startlink[1]{}%
\providecommand \@@endlink[0]{}%
\providecommand \url  [0]{\begingroup\@sanitize@url \@url }%
\providecommand \@url [1]{\endgroup\@href {#1}{\urlprefix }}%
\providecommand \urlprefix  [0]{URL }%
\providecommand \Eprint [0]{\href }%
\providecommand \doibase [0]{https://doi.org/}%
\providecommand \selectlanguage [0]{\@gobble}%
\providecommand \bibinfo  [0]{\@secondoftwo}%
\providecommand \bibfield  [0]{\@secondoftwo}%
\providecommand \translation [1]{[#1]}%
\providecommand \BibitemOpen [0]{}%
\providecommand \bibitemStop [0]{}%
\providecommand \bibitemNoStop [0]{.\EOS\space}%
\providecommand \EOS [0]{\spacefactor3000\relax}%
\providecommand \BibitemShut  [1]{\csname bibitem#1\endcsname}%
\let\auto@bib@innerbib\@empty
\bibitem [{\citenamefont {Paris}\ and\ \citenamefont
  {\v{R}eh\'a\v{c}ek}(2004)}]{lnp:2004uq}%
  \BibitemOpen
  \bibinfo {editor} {\bibfnamefont {M.~G.~A.}\ \bibnamefont {Paris}}\ and\
  \bibinfo {editor} {\bibfnamefont {J.}~\bibnamefont {\v{R}eh\'a\v{c}ek}},\
  eds.,\ \href@noop {} {\emph {\bibinfo {title} {Quantum State Estimation}}},\
  \bibinfo {series} {Lect. Not. Phys.}, Vol.\ \bibinfo {volume} {649}\
  (\bibinfo  {publisher} {Springer},\ \bibinfo {address} {Berlin},\ \bibinfo
  {year} {2004})\BibitemShut {NoStop}%
\bibitem [{\citenamefont {O'Donnell}\ and\ \citenamefont
  {Wright}(2016)}]{ODonnell:2016wk}%
  \BibitemOpen
  \bibfield  {author} {\bibinfo {author} {\bibfnamefont {R.}~\bibnamefont
  {O'Donnell}}\ and\ \bibinfo {author} {\bibfnamefont {J.}~\bibnamefont
  {Wright}},\ }\bibfield  {title} {\bibinfo {title} {Efficient quantum
  tomography},\ }in\ \href {https://doi.org/10.1145/2897518.2897544} {\emph
  {\bibinfo {booktitle} {Proc. forty-eighth annual ACM symposium on Theory of
  Computing}}}\ (\bibinfo  {publisher} {Association for Computing Machinery},\
  \bibinfo {address} {Cambridge, MA, USA},\ \bibinfo {year} {2016})\ pp.\
  \bibinfo {pages} {899--912}\BibitemShut {NoStop}%
\bibitem [{\citenamefont {Artiles}\ \emph {et~al.}(2005)\citenamefont
  {Artiles}, \citenamefont {Gill},\ and\ \citenamefont {Gut¸{\u
  a}}}]{Artiles:2005wk}%
  \BibitemOpen
  \bibfield  {author} {\bibinfo {author} {\bibfnamefont {L.~M.}\ \bibnamefont
  {Artiles}}, \bibinfo {author} {\bibfnamefont {R.~D.}\ \bibnamefont {Gill}},\
  and\ \bibinfo {author} {\bibfnamefont {M.~I.}\ \bibnamefont {Gut¸{\u a}}},\
  }\bibfield  {title} {\bibinfo {title} {An invitation to quantum tomography},\
  }\href {https://doi.org/https://doi.org/10.1111/j.1467-9868.2005.00491.x}
  {\bibfield  {journal} {\bibinfo  {journal} {J. Roy. Stat. Soc. B}\ }\textbf
  {\bibinfo {volume} {67}},\ \bibinfo {pages} {109} (\bibinfo {year}
  {2005})}\BibitemShut {NoStop}%
\bibitem [{\citenamefont {Tarantola}(2005)}]{Tarantola:2005aa}%
  \BibitemOpen
  \bibfield  {author} {\bibinfo {author} {\bibfnamefont {A.}~\bibnamefont
  {Tarantola}},\ }\href {https://doi.org/doi:10.1137/1.9780898717921} {\emph
  {\bibinfo {title} {Inverse {P}roblem {T}heory and {M}ethods for {M}odel
  {P}arameter {E}stimation}}}\ (\bibinfo  {publisher} {SIAM},\ \bibinfo
  {address} {Philadelphia},\ \bibinfo {year} {2005})\ p.\ \bibinfo {pages}
  {348}\BibitemShut {NoStop}%
\bibitem [{\citenamefont {Aster}\ \emph {et~al.}(2013)\citenamefont {Aster},
  \citenamefont {Borchers},\ and\ \citenamefont {Thurber}}]{Aster:2013aa}%
  \BibitemOpen
  \bibfield  {author} {\bibinfo {author} {\bibfnamefont {R.}~\bibnamefont
  {Aster}}, \bibinfo {author} {\bibfnamefont {B.}~\bibnamefont {Borchers}},\
  and\ \bibinfo {author} {\bibfnamefont {C.}~\bibnamefont {Thurber}},\
  }\href@noop {} {\emph {\bibinfo {title} {Parameter Estimation and Inverse
  Problems}}},\ \bibinfo {edition} {2nd}\ ed.\ (\bibinfo  {publisher}
  {Academic},\ \bibinfo {address} {Boston},\ \bibinfo {year}
  {2013})\BibitemShut {NoStop}%
\bibitem [{\citenamefont {Qi}\ \emph {et~al.}(2013)\citenamefont {Qi},
  \citenamefont {Hou}, \citenamefont {Li}, \citenamefont {Dong}, \citenamefont
  {Xiang},\ and\ \citenamefont {Guo}}]{Qi:2013ts}%
  \BibitemOpen
  \bibfield  {author} {\bibinfo {author} {\bibfnamefont {B.}~\bibnamefont
  {Qi}}, \bibinfo {author} {\bibfnamefont {Z.}~\bibnamefont {Hou}}, \bibinfo
  {author} {\bibfnamefont {L.}~\bibnamefont {Li}}, \bibinfo {author}
  {\bibfnamefont {D.}~\bibnamefont {Dong}}, \bibinfo {author} {\bibfnamefont
  {G.}~\bibnamefont {Xiang}},\ and\ \bibinfo {author} {\bibfnamefont
  {G.}~\bibnamefont {Guo}},\ }\bibfield  {title} {\bibinfo {title} {Quantum
  state tomography via linear regression estimation},\ }\href
  {https://doi.org/10.1038/srep03496} {\bibfield  {journal} {\bibinfo
  {journal} {Sci. Rep.}\ }\textbf {\bibinfo {volume} {3}},\ \bibinfo {pages}
  {3496} (\bibinfo {year} {2013})}\BibitemShut {NoStop}%
\bibitem [{\citenamefont {Blume-Kohout}(2010)}]{Blume-Kohout:2010uf}%
  \BibitemOpen
  \bibfield  {author} {\bibinfo {author} {\bibfnamefont {R.}~\bibnamefont
  {Blume-Kohout}},\ }\bibfield  {title} {\bibinfo {title} {Optimal, reliable
  estimation of quantum states},\ }\href
  {https://doi.org/10.1088/1367-2630/12/4/043034} {\bibfield  {journal}
  {\bibinfo  {journal} {New J. Phys.}\ }\textbf {\bibinfo {volume} {12}},\
  \bibinfo {pages} {043034} (\bibinfo {year} {2010})}\BibitemShut {NoStop}%
\bibitem [{\citenamefont {Granade}\ \emph {et~al.}(2016)\citenamefont
  {Granade}, \citenamefont {Combes},\ and\ \citenamefont
  {Cory}}]{Granade:2016tr}%
  \BibitemOpen
  \bibfield  {author} {\bibinfo {author} {\bibfnamefont {C.}~\bibnamefont
  {Granade}}, \bibinfo {author} {\bibfnamefont {J.}~\bibnamefont {Combes}},\
  and\ \bibinfo {author} {\bibfnamefont {D.~G.}\ \bibnamefont {Cory}},\
  }\bibfield  {title} {\bibinfo {title} {Practical bayesian tomography},\
  }\href {https://doi.org/10.1088/1367-2630/18/3/033024} {\bibfield  {journal}
  {\bibinfo  {journal} {New J. Phys.}\ }\textbf {\bibinfo {volume} {18}},\
  \bibinfo {pages} {033024} (\bibinfo {year} {2016})}\BibitemShut {NoStop}%
\bibitem [{\citenamefont {Gross}\ \emph {et~al.}(2010)\citenamefont {Gross},
  \citenamefont {Liu}, \citenamefont {Flammia}, \citenamefont {Becker},\ and\
  \citenamefont {Eisert}}]{Gross:2010aa}%
  \BibitemOpen
  \bibfield  {author} {\bibinfo {author} {\bibfnamefont {D.}~\bibnamefont
  {Gross}}, \bibinfo {author} {\bibfnamefont {Y.-K.}\ \bibnamefont {Liu}},
  \bibinfo {author} {\bibfnamefont {S.~T.}\ \bibnamefont {Flammia}}, \bibinfo
  {author} {\bibfnamefont {S.}~\bibnamefont {Becker}},\ and\ \bibinfo {author}
  {\bibfnamefont {J.}~\bibnamefont {Eisert}},\ }\bibfield  {title} {\bibinfo
  {title} {Quantum state tomography via compressed sensing},\ }\href
  {http://link.aps.org/doi/10.1103/PhysRevLett.105.150401} {\bibfield
  {journal} {\bibinfo  {journal} {Phys. Rev. Lett.}\ }\textbf {\bibinfo
  {volume} {105}},\ \bibinfo {pages} {150401} (\bibinfo {year}
  {2010})}\BibitemShut {NoStop}%
\bibitem [{\citenamefont {Ahn}\ \emph {et~al.}(2019)\citenamefont {Ahn},
  \citenamefont {Teo}, \citenamefont {Jeong}, \citenamefont {Bouchard},
  \citenamefont {Hufnagel}, \citenamefont {Karimi}, \citenamefont {Koutn{\'y}},
  \citenamefont {{\v R}eh{\'a}{\v c}ek}, \citenamefont {Hradil}, \citenamefont
  {Leuchs},\ and\ \citenamefont {S{\'a}nchez-Soto}}]{Ahn:2019uw}%
  \BibitemOpen
  \bibfield  {author} {\bibinfo {author} {\bibfnamefont {D.}~\bibnamefont
  {Ahn}}, \bibinfo {author} {\bibfnamefont {Y.~S.}\ \bibnamefont {Teo}},
  \bibinfo {author} {\bibfnamefont {H.}~\bibnamefont {Jeong}}, \bibinfo
  {author} {\bibfnamefont {F.}~\bibnamefont {Bouchard}}, \bibinfo {author}
  {\bibfnamefont {F.}~\bibnamefont {Hufnagel}}, \bibinfo {author}
  {\bibfnamefont {E.}~\bibnamefont {Karimi}}, \bibinfo {author} {\bibfnamefont
  {D.}~\bibnamefont {Koutn{\'y}}}, \bibinfo {author} {\bibfnamefont
  {J.}~\bibnamefont {{\v R}eh{\'a}{\v c}ek}}, \bibinfo {author} {\bibfnamefont
  {Z.}~\bibnamefont {Hradil}}, \bibinfo {author} {\bibfnamefont
  {G.}~\bibnamefont {Leuchs}},\ and\ \bibinfo {author} {\bibfnamefont {L.~L.}\
  \bibnamefont {S{\'a}nchez-Soto}},\ }\bibfield  {title} {\bibinfo {title}
  {Adaptive compressive tomography with no a priori information},\ }\href
  {https://doi.org/10.1103/PhysRevLett.122.100404} {\bibfield  {journal}
  {\bibinfo  {journal} {Phys. Rev. Lett.}\ }\textbf {\bibinfo {volume} {122}},\
  \bibinfo {pages} {100404} (\bibinfo {year} {2019})}\BibitemShut {NoStop}%
\bibitem [{\citenamefont {Cramer}\ \emph {et~al.}(2010)\citenamefont {Cramer},
  \citenamefont {Plenio}, \citenamefont {Flammia}, \citenamefont {Somma},
  \citenamefont {Gross}, \citenamefont {Bartlett}, \citenamefont
  {Landon-Cardinal}, \citenamefont {Poulin},\ and\ \citenamefont
  {Liu}}]{Cramer:2010aa}%
  \BibitemOpen
  \bibfield  {author} {\bibinfo {author} {\bibfnamefont {M.}~\bibnamefont
  {Cramer}}, \bibinfo {author} {\bibfnamefont {M.~B.}\ \bibnamefont {Plenio}},
  \bibinfo {author} {\bibfnamefont {S.~T.}\ \bibnamefont {Flammia}}, \bibinfo
  {author} {\bibfnamefont {R.}~\bibnamefont {Somma}}, \bibinfo {author}
  {\bibfnamefont {D.}~\bibnamefont {Gross}}, \bibinfo {author} {\bibfnamefont
  {S.~D.}\ \bibnamefont {Bartlett}}, \bibinfo {author} {\bibfnamefont
  {O.}~\bibnamefont {Landon-Cardinal}}, \bibinfo {author} {\bibfnamefont
  {D.}~\bibnamefont {Poulin}},\ and\ \bibinfo {author} {\bibfnamefont {Y.~K.}\
  \bibnamefont {Liu}},\ }\bibfield  {title} {\bibinfo {title} {Efficient
  quantum state tomography},\ }\href {http://dx.doi.org/10.1038/ncomms1147}
  {\bibfield  {journal} {\bibinfo  {journal} {Nat. Commun.}\ }\textbf {\bibinfo
  {volume} {1}},\ \bibinfo {pages} {149 EP} (\bibinfo {year}
  {2010})}\BibitemShut {NoStop}%
\bibitem [{\citenamefont {Lanyon}\ \emph {et~al.}(2017)\citenamefont {Lanyon},
  \citenamefont {Maier}, \citenamefont {Holz{\"a}pfel}, \citenamefont
  {Baumgratz}, \citenamefont {Hempel}, \citenamefont {Jurcevic}, \citenamefont
  {Dhand}, \citenamefont {Buyskikh}, \citenamefont {Daley}, \citenamefont
  {Cramer}, \citenamefont {Plenio}, \citenamefont {Blatt},\ and\ \citenamefont
  {Roos}}]{Lanyon:2017vi}%
  \BibitemOpen
  \bibfield  {author} {\bibinfo {author} {\bibfnamefont {B.~P.}\ \bibnamefont
  {Lanyon}}, \bibinfo {author} {\bibfnamefont {C.}~\bibnamefont {Maier}},
  \bibinfo {author} {\bibfnamefont {M.}~\bibnamefont {Holz{\"a}pfel}}, \bibinfo
  {author} {\bibfnamefont {T.}~\bibnamefont {Baumgratz}}, \bibinfo {author}
  {\bibfnamefont {C.}~\bibnamefont {Hempel}}, \bibinfo {author} {\bibfnamefont
  {P.}~\bibnamefont {Jurcevic}}, \bibinfo {author} {\bibfnamefont
  {I.}~\bibnamefont {Dhand}}, \bibinfo {author} {\bibfnamefont {A.~S.}\
  \bibnamefont {Buyskikh}}, \bibinfo {author} {\bibfnamefont {A.~J.}\
  \bibnamefont {Daley}}, \bibinfo {author} {\bibfnamefont {M.}~\bibnamefont
  {Cramer}}, \bibinfo {author} {\bibfnamefont {M.~B.}\ \bibnamefont {Plenio}},
  \bibinfo {author} {\bibfnamefont {R.}~\bibnamefont {Blatt}},\ and\ \bibinfo
  {author} {\bibfnamefont {C.~F.}\ \bibnamefont {Roos}},\ }\bibfield  {title}
  {\bibinfo {title} {Efficient tomography of a quantum many-body system},\
  }\href {https://doi.org/10.1038/nphys4244} {\bibfield  {journal} {\bibinfo
  {journal} {Nat. Phys.}\ }\textbf {\bibinfo {volume} {13}},\ \bibinfo {pages}
  {1158} (\bibinfo {year} {2017})}\BibitemShut {NoStop}%
\bibitem [{\citenamefont {Hradil}(1997)}]{Hradil:1997tt}%
  \BibitemOpen
  \bibfield  {author} {\bibinfo {author} {\bibfnamefont {Z.}~\bibnamefont
  {Hradil}},\ }\bibfield  {title} {\bibinfo {title} {Quantum-state
  estimation},\ }\href {https://doi.org/10.1103/PhysRevA.55.R1561} {\bibfield
  {journal} {\bibinfo  {journal} {Phys. Rev. A}\ }\textbf {\bibinfo {volume}
  {55}},\ \bibinfo {pages} {R1561} (\bibinfo {year} {1997})}\BibitemShut
  {NoStop}%
\bibitem [{\citenamefont {James}\ \emph {et~al.}(2001)\citenamefont {James},
  \citenamefont {Kwiat}, \citenamefont {Munro},\ and\ \citenamefont
  {White}}]{James:2001ut}%
  \BibitemOpen
  \bibfield  {author} {\bibinfo {author} {\bibfnamefont {D.~F.~V.}\
  \bibnamefont {James}}, \bibinfo {author} {\bibfnamefont {P.~G.}\ \bibnamefont
  {Kwiat}}, \bibinfo {author} {\bibfnamefont {W.~J.}\ \bibnamefont {Munro}},\
  and\ \bibinfo {author} {\bibfnamefont {A.~G.}\ \bibnamefont {White}},\
  }\bibfield  {title} {\bibinfo {title} {Measurement of qubits},\ }\href
  {https://doi.org/10.1103/PhysRevA.64.052312} {\bibfield  {journal} {\bibinfo
  {journal} {Phys. Rev. A}\ }\textbf {\bibinfo {volume} {64}},\ \bibinfo
  {pages} {052312} (\bibinfo {year} {2001})}\BibitemShut {NoStop}%
\bibitem [{\citenamefont {Murphy}(2012)}]{Murphy:2012uq}%
  \BibitemOpen
  \bibfield  {author} {\bibinfo {author} {\bibfnamefont {K.}~\bibnamefont
  {Murphy}},\ }\href@noop {} {\emph {\bibinfo {title} {Machine Learning: A
  Probabilistic Perspective}}}\ (\bibinfo  {publisher} {The MIT Press},\
  \bibinfo {address} {Cambridge},\ \bibinfo {year} {2012})\BibitemShut
  {NoStop}%
\bibitem [{\citenamefont {Nielsen}(2018)}]{Nielsen:2018ul}%
  \BibitemOpen
  \bibfield  {author} {\bibinfo {author} {\bibfnamefont {M.~A.}\ \bibnamefont
  {Nielsen}},\ }\href {http://neuralnetworksanddeeplearning.com/} {\bibinfo
  {title} {Neural networks and deep learning}} (\bibinfo {year}
  {2018})\BibitemShut {NoStop}%
\bibitem [{\citenamefont {Mehta}\ \emph {et~al.}(2019)\citenamefont {Mehta},
  \citenamefont {Bukov}, \citenamefont {Wang}, \citenamefont {Day},
  \citenamefont {Richardson}, \citenamefont {Fisher},\ and\ \citenamefont
  {Schwab}}]{Mehta:2019tt}%
  \BibitemOpen
  \bibfield  {author} {\bibinfo {author} {\bibfnamefont {P.}~\bibnamefont
  {Mehta}}, \bibinfo {author} {\bibfnamefont {M.}~\bibnamefont {Bukov}},
  \bibinfo {author} {\bibfnamefont {C.-H.}\ \bibnamefont {Wang}}, \bibinfo
  {author} {\bibfnamefont {A.~G.~R.}\ \bibnamefont {Day}}, \bibinfo {author}
  {\bibfnamefont {C.}~\bibnamefont {Richardson}}, \bibinfo {author}
  {\bibfnamefont {C.~K.}\ \bibnamefont {Fisher}},\ and\ \bibinfo {author}
  {\bibfnamefont {D.~J.}\ \bibnamefont {Schwab}},\ }\bibfield  {title}
  {\bibinfo {title} {A high-bias, low-variance introduction to machine learning
  for physicists},\ }\href
  {https://doi.org/https://doi.org/10.1016/j.physrep.2019.03.001} {\bibfield
  {journal} {\bibinfo  {journal} {Phys. Rep.}\ }\textbf {\bibinfo {volume}
  {810}},\ \bibinfo {pages} {1} (\bibinfo {year} {2019})}\BibitemShut {NoStop}%
\bibitem [{\citenamefont {Rem}\ \emph {et~al.}(2019)\citenamefont {Rem},
  \citenamefont {K{\"a}ming}, \citenamefont {Tarnowski}, \citenamefont
  {Asteria}, \citenamefont {Fl{\"a}schner}, \citenamefont {Becker},
  \citenamefont {Sengstock},\ and\ \citenamefont {Weitenberg}}]{Rem:2019vc}%
  \BibitemOpen
  \bibfield  {author} {\bibinfo {author} {\bibfnamefont {B.~S.}\ \bibnamefont
  {Rem}}, \bibinfo {author} {\bibfnamefont {N.}~\bibnamefont {K{\"a}ming}},
  \bibinfo {author} {\bibfnamefont {M.}~\bibnamefont {Tarnowski}}, \bibinfo
  {author} {\bibfnamefont {L.}~\bibnamefont {Asteria}}, \bibinfo {author}
  {\bibfnamefont {N.}~\bibnamefont {Fl{\"a}schner}}, \bibinfo {author}
  {\bibfnamefont {C.}~\bibnamefont {Becker}}, \bibinfo {author} {\bibfnamefont
  {K.}~\bibnamefont {Sengstock}},\ and\ \bibinfo {author} {\bibfnamefont
  {C.}~\bibnamefont {Weitenberg}},\ }\bibfield  {title} {\bibinfo {title}
  {Identifying quantum phase transitions using artificial neural networks on
  experimental data},\ }\href {https://doi.org/10.1038/s41567-019-0554-0}
  {\bibfield  {journal} {\bibinfo  {journal} {Nat. Phys.}\ }\textbf {\bibinfo
  {volume} {15}},\ \bibinfo {pages} {917} (\bibinfo {year} {2019})}\BibitemShut
  {NoStop}%
\bibitem [{\citenamefont {Gebhart}\ and\ \citenamefont
  {Bohmann}(2020)}]{Gebhart:2020tx}%
  \BibitemOpen
  \bibfield  {author} {\bibinfo {author} {\bibfnamefont {V.}~\bibnamefont
  {Gebhart}}\ and\ \bibinfo {author} {\bibfnamefont {M.}~\bibnamefont
  {Bohmann}},\ }\bibfield  {title} {\bibinfo {title} {Neural-network approach
  for identifying nonclassicality from click-counting data},\ }\href
  {https://doi.org/10.1103/PhysRevResearch.2.023150} {\bibfield  {journal}
  {\bibinfo  {journal} {Phys. Rev. Research}\ }\textbf {\bibinfo {volume}
  {2}},\ \bibinfo {pages} {023150} (\bibinfo {year} {2020})}\BibitemShut
  {NoStop}%
\bibitem [{\citenamefont {You}\ \emph {et~al.}(2020)\citenamefont {You},
  \citenamefont {Quiroz-Ju{\'a}rez}, \citenamefont {Lambert}, \citenamefont
  {Bhusal}, \citenamefont {Dong}, \citenamefont {Perez-Leija}, \citenamefont
  {Javaid}, \citenamefont {Le{\'o}n-Montiel},\ and\ \citenamefont
  {Maga{\~n}a-Loaiza}}]{You:2020uj}%
  \BibitemOpen
  \bibfield  {author} {\bibinfo {author} {\bibfnamefont {C.}~\bibnamefont
  {You}}, \bibinfo {author} {\bibfnamefont {M.~A.}\ \bibnamefont
  {Quiroz-Ju{\'a}rez}}, \bibinfo {author} {\bibfnamefont {A.}~\bibnamefont
  {Lambert}}, \bibinfo {author} {\bibfnamefont {N.}~\bibnamefont {Bhusal}},
  \bibinfo {author} {\bibfnamefont {C.}~\bibnamefont {Dong}}, \bibinfo {author}
  {\bibfnamefont {A.}~\bibnamefont {Perez-Leija}}, \bibinfo {author}
  {\bibfnamefont {A.}~\bibnamefont {Javaid}}, \bibinfo {author} {\bibfnamefont
  {R.~J.}\ \bibnamefont {Le{\'o}n-Montiel}},\ and\ \bibinfo {author}
  {\bibfnamefont {O.~S.}\ \bibnamefont {Maga{\~n}a-Loaiza}},\ }\bibfield
  {title} {\bibinfo {title} {Identification of light sources using machine
  learning},\ }\href {https://doi.org/10.1063/1.5133846} {\bibfield  {journal}
  {\bibinfo  {journal} {Appl. Phys. Rev.}\ }\textbf {\bibinfo {volume} {7}},\
  \bibinfo {pages} {021404} (\bibinfo {year} {2020})}\BibitemShut {NoStop}%
\bibitem [{\citenamefont {Harney}\ \emph {et~al.}(2020)\citenamefont {Harney},
  \citenamefont {Pirandola}, \citenamefont {Ferraro},\ and\ \citenamefont
  {Paternostro}}]{Harney:2020ta}%
  \BibitemOpen
  \bibfield  {author} {\bibinfo {author} {\bibfnamefont {C.}~\bibnamefont
  {Harney}}, \bibinfo {author} {\bibfnamefont {S.}~\bibnamefont {Pirandola}},
  \bibinfo {author} {\bibfnamefont {A.}~\bibnamefont {Ferraro}},\ and\ \bibinfo
  {author} {\bibfnamefont {M.}~\bibnamefont {Paternostro}},\ }\bibfield
  {title} {\bibinfo {title} {Entanglement classification via neural network
  quantum states},\ }\href {https://doi.org/10.1088/1367-2630/ab783d}
  {\bibfield  {journal} {\bibinfo  {journal} {New J. Phys.}\ }\textbf {\bibinfo
  {volume} {22}},\ \bibinfo {pages} {045001} (\bibinfo {year}
  {2020})}\BibitemShut {NoStop}%
\bibitem [{\citenamefont {Torlai}\ and\ \citenamefont
  {Melko}(2017)}]{Torlai:2017tz}%
  \BibitemOpen
  \bibfield  {author} {\bibinfo {author} {\bibfnamefont {G.}~\bibnamefont
  {Torlai}}\ and\ \bibinfo {author} {\bibfnamefont {R.~G.}\ \bibnamefont
  {Melko}},\ }\bibfield  {title} {\bibinfo {title} {Neural decoder for
  topological codes},\ }\href {https://doi.org/10.1103/PhysRevLett.119.030501}
  {\bibfield  {journal} {\bibinfo  {journal} {Phys. Rev. Lett.}\ }\textbf
  {\bibinfo {volume} {119}},\ \bibinfo {pages} {030501} (\bibinfo {year}
  {2017})}\BibitemShut {NoStop}%
\bibitem [{\citenamefont {Baireuther}\ \emph {et~al.}(2018)\citenamefont
  {Baireuther}, \citenamefont {O'Brien}, \citenamefont {Tarasinski},\ and\
  \citenamefont {Beenakker}}]{Baireuther:2018tj}%
  \BibitemOpen
  \bibfield  {author} {\bibinfo {author} {\bibfnamefont {P.}~\bibnamefont
  {Baireuther}}, \bibinfo {author} {\bibfnamefont {T.~E.}\ \bibnamefont
  {O'Brien}}, \bibinfo {author} {\bibfnamefont {B.}~\bibnamefont
  {Tarasinski}},\ and\ \bibinfo {author} {\bibfnamefont {C.~W.~J.}\
  \bibnamefont {Beenakker}},\ }\bibfield  {title} {\bibinfo {title}
  {Machine-learning-assisted correction of correlated qubit errors in a
  topological code},\ }\href {https://doi.org/10.22331/q-2018-01-29-48}
  {\bibfield  {journal} {\bibinfo  {journal} {{Quantum}}\ }\textbf {\bibinfo
  {volume} {2}},\ \bibinfo {pages} {48} (\bibinfo {year} {2018})}\BibitemShut
  {NoStop}%
\bibitem [{\citenamefont {Krastanov}\ and\ \citenamefont
  {Jiang}(2017)}]{Krastanov:2017vr}%
  \BibitemOpen
  \bibfield  {author} {\bibinfo {author} {\bibfnamefont {S.}~\bibnamefont
  {Krastanov}}\ and\ \bibinfo {author} {\bibfnamefont {L.}~\bibnamefont
  {Jiang}},\ }\bibfield  {title} {\bibinfo {title} {Deep neural network
  probabilistic decoder for stabilizer codes},\ }\href
  {https://doi.org/10.1038/s41598-017-11266-1} {\bibfield  {journal} {\bibinfo
  {journal} {Sci. Rep.}\ }\textbf {\bibinfo {volume} {7}},\ \bibinfo {pages}
  {11003} (\bibinfo {year} {2017})}\BibitemShut {NoStop}%
\bibitem [{\citenamefont {Fitzek}\ \emph {et~al.}(2020)\citenamefont {Fitzek},
  \citenamefont {Eliasson}, \citenamefont {Kockum},\ and\ \citenamefont
  {Granath}}]{Fitzek:2020ux}%
  \BibitemOpen
  \bibfield  {author} {\bibinfo {author} {\bibfnamefont {D.}~\bibnamefont
  {Fitzek}}, \bibinfo {author} {\bibfnamefont {M.}~\bibnamefont {Eliasson}},
  \bibinfo {author} {\bibfnamefont {A.~F.}\ \bibnamefont {Kockum}},\ and\
  \bibinfo {author} {\bibfnamefont {M.}~\bibnamefont {Granath}},\ }\bibfield
  {title} {\bibinfo {title} {Deep q-learning decoder for depolarizing noise on
  the toric code},\ }\href {https://doi.org/10.1103/PhysRevResearch.2.023230}
  {\bibfield  {journal} {\bibinfo  {journal} {Phys. Rev. Research}\ }\textbf
  {\bibinfo {volume} {2}},\ \bibinfo {pages} {023230} (\bibinfo {year}
  {2020})}\BibitemShut {NoStop}%
\bibitem [{\citenamefont {Flurin}\ \emph {et~al.}(2020)\citenamefont {Flurin},
  \citenamefont {Martin}, \citenamefont {Hacohen-Gourgy},\ and\ \citenamefont
  {Siddiqi}}]{Flurin:2020vf}%
  \BibitemOpen
  \bibfield  {author} {\bibinfo {author} {\bibfnamefont {E.}~\bibnamefont
  {Flurin}}, \bibinfo {author} {\bibfnamefont {L.~S.}\ \bibnamefont {Martin}},
  \bibinfo {author} {\bibfnamefont {S.}~\bibnamefont {Hacohen-Gourgy}},\ and\
  \bibinfo {author} {\bibfnamefont {I.}~\bibnamefont {Siddiqi}},\ }\bibfield
  {title} {\bibinfo {title} {Using a recurrent neural network to reconstruct
  quantum dynamics of a superconducting qubit from physical observations},\
  }\href {https://doi.org/10.1103/PhysRevX.10.011006} {\bibfield  {journal}
  {\bibinfo  {journal} {Phys. Rev. X}\ }\textbf {\bibinfo {volume} {10}},\
  \bibinfo {pages} {011006} (\bibinfo {year} {2020})}\BibitemShut {NoStop}%
\bibitem [{\citenamefont {Wittler}\ \emph {et~al.}(2021)\citenamefont
  {Wittler}, \citenamefont {Roy}, \citenamefont {Pack}, \citenamefont
  {Werninghaus}, \citenamefont {Roy}, \citenamefont {Egger}, \citenamefont
  {Filipp}, \citenamefont {Wilhelm},\ and\ \citenamefont
  {Machnes}}]{Wittler:2021vj}%
  \BibitemOpen
  \bibfield  {author} {\bibinfo {author} {\bibfnamefont {N.}~\bibnamefont
  {Wittler}}, \bibinfo {author} {\bibfnamefont {F.}~\bibnamefont {Roy}},
  \bibinfo {author} {\bibfnamefont {K.}~\bibnamefont {Pack}}, \bibinfo {author}
  {\bibfnamefont {M.}~\bibnamefont {Werninghaus}}, \bibinfo {author}
  {\bibfnamefont {A.~S.}\ \bibnamefont {Roy}}, \bibinfo {author} {\bibfnamefont
  {D.~J.}\ \bibnamefont {Egger}}, \bibinfo {author} {\bibfnamefont
  {S.}~\bibnamefont {Filipp}}, \bibinfo {author} {\bibfnamefont {F.~K.}\
  \bibnamefont {Wilhelm}},\ and\ \bibinfo {author} {\bibfnamefont
  {S.}~\bibnamefont {Machnes}},\ }\bibfield  {title} {\bibinfo {title}
  {Integrated tool set for control, calibration, and characterization of
  quantum devices applied to superconducting qubits},\ }\href
  {https://doi.org/10.1103/PhysRevApplied.15.034080} {\bibfield  {journal}
  {\bibinfo  {journal} {Phys. Rev. Applied}\ }\textbf {\bibinfo {volume}
  {15}},\ \bibinfo {pages} {034080} (\bibinfo {year} {2021})}\BibitemShut
  {NoStop}%
\bibitem [{\citenamefont {Leung}\ \emph {et~al.}(2017)\citenamefont {Leung},
  \citenamefont {Abdelhafez}, \citenamefont {Koch},\ and\ \citenamefont
  {Schuster}}]{Leung:2017tx}%
  \BibitemOpen
  \bibfield  {author} {\bibinfo {author} {\bibfnamefont {N.}~\bibnamefont
  {Leung}}, \bibinfo {author} {\bibfnamefont {M.}~\bibnamefont {Abdelhafez}},
  \bibinfo {author} {\bibfnamefont {J.}~\bibnamefont {Koch}},\ and\ \bibinfo
  {author} {\bibfnamefont {D.}~\bibnamefont {Schuster}},\ }\bibfield  {title}
  {\bibinfo {title} {Speedup for quantum optimal control from automatic
  differentiation based on graphics processing units},\ }\href
  {https://doi.org/10.1103/PhysRevA.95.042318} {\bibfield  {journal} {\bibinfo
  {journal} {Phys. Rev. A}\ }\textbf {\bibinfo {volume} {95}},\ \bibinfo
  {pages} {042318} (\bibinfo {year} {2017})}\BibitemShut {NoStop}%
\bibitem [{\citenamefont {Krenn}\ \emph {et~al.}(2016)\citenamefont {Krenn},
  \citenamefont {Malik}, \citenamefont {Fickler}, \citenamefont {Lapkiewicz},\
  and\ \citenamefont {Zeilinger}}]{Krenn:2016te}%
  \BibitemOpen
  \bibfield  {author} {\bibinfo {author} {\bibfnamefont {M.}~\bibnamefont
  {Krenn}}, \bibinfo {author} {\bibfnamefont {M.}~\bibnamefont {Malik}},
  \bibinfo {author} {\bibfnamefont {R.}~\bibnamefont {Fickler}}, \bibinfo
  {author} {\bibfnamefont {R.}~\bibnamefont {Lapkiewicz}},\ and\ \bibinfo
  {author} {\bibfnamefont {A.}~\bibnamefont {Zeilinger}},\ }\bibfield  {title}
  {\bibinfo {title} {Automated search for new quantum experiments},\ }\href
  {https://doi.org/10.1103/PhysRevLett.116.090405} {\bibfield  {journal}
  {\bibinfo  {journal} {Phys. Rev. Lett.}\ }\textbf {\bibinfo {volume} {116}},\
  \bibinfo {pages} {090405} (\bibinfo {year} {2016})}\BibitemShut {NoStop}%
\bibitem [{\citenamefont {Melnikov}\ \emph {et~al.}(2018)\citenamefont
  {Melnikov}, \citenamefont {Poulsen~Nautrup}, \citenamefont {Krenn},
  \citenamefont {Dunjko}, \citenamefont {Tiersch}, \citenamefont {Zeilinger},\
  and\ \citenamefont {Briegel}}]{Melnikov:2018tv}%
  \BibitemOpen
  \bibfield  {author} {\bibinfo {author} {\bibfnamefont {A.~A.}\ \bibnamefont
  {Melnikov}}, \bibinfo {author} {\bibfnamefont {H.}~\bibnamefont
  {Poulsen~Nautrup}}, \bibinfo {author} {\bibfnamefont {M.}~\bibnamefont
  {Krenn}}, \bibinfo {author} {\bibfnamefont {V.}~\bibnamefont {Dunjko}},
  \bibinfo {author} {\bibfnamefont {M.}~\bibnamefont {Tiersch}}, \bibinfo
  {author} {\bibfnamefont {A.}~\bibnamefont {Zeilinger}},\ and\ \bibinfo
  {author} {\bibfnamefont {H.~J.}\ \bibnamefont {Briegel}},\ }\bibfield
  {title} {\bibinfo {title} {Active learning machine learns to create new
  quantum experiments},\ }\href
  {https://www.pnas.org/content/pnas/115/6/1221.full.pdf} {\bibfield  {journal}
  {\bibinfo  {journal} {Proc. Natl. Acad. Sci. USA}\ }\textbf {\bibinfo
  {volume} {115}},\ \bibinfo {pages} {1221} (\bibinfo {year}
  {2018})}\BibitemShut {NoStop}%
\bibitem [{\citenamefont {O'Driscoll}\ \emph {et~al.}(2019)\citenamefont
  {O'Driscoll}, \citenamefont {Nichols},\ and\ \citenamefont
  {Knott}}]{ODriscoll:2019wz}%
  \BibitemOpen
  \bibfield  {author} {\bibinfo {author} {\bibfnamefont {L.}~\bibnamefont
  {O'Driscoll}}, \bibinfo {author} {\bibfnamefont {R.}~\bibnamefont
  {Nichols}},\ and\ \bibinfo {author} {\bibfnamefont {P.~A.}\ \bibnamefont
  {Knott}},\ }\bibfield  {title} {\bibinfo {title} {A hybrid machine learning
  algorithm for designing quantum experiments},\ }\href
  {https://doi.org/10.1007/s42484-019-00003-8} {\bibfield  {journal} {\bibinfo
  {journal} {Quantum Machine Intelligence}\ }\textbf {\bibinfo {volume} {1}},\
  \bibinfo {pages} {5} (\bibinfo {year} {2019})}\BibitemShut {NoStop}%
\bibitem [{\citenamefont {Torlai}\ \emph {et~al.}(2018)\citenamefont {Torlai},
  \citenamefont {Mazzola}, \citenamefont {Carrasquilla}, \citenamefont
  {Troyer}, \citenamefont {Melko},\ and\ \citenamefont
  {Carleo}}]{Torlai:2018wn}%
  \BibitemOpen
  \bibfield  {author} {\bibinfo {author} {\bibfnamefont {G.}~\bibnamefont
  {Torlai}}, \bibinfo {author} {\bibfnamefont {G.}~\bibnamefont {Mazzola}},
  \bibinfo {author} {\bibfnamefont {J.}~\bibnamefont {Carrasquilla}}, \bibinfo
  {author} {\bibfnamefont {M.}~\bibnamefont {Troyer}}, \bibinfo {author}
  {\bibfnamefont {R.}~\bibnamefont {Melko}},\ and\ \bibinfo {author}
  {\bibfnamefont {G.}~\bibnamefont {Carleo}},\ }\bibfield  {title} {\bibinfo
  {title} {Neural-network quantum state tomography},\ }\href
  {https://doi.org/10.1038/s41567-018-0048-5} {\bibfield  {journal} {\bibinfo
  {journal} {Nat. Phys.}\ }\textbf {\bibinfo {volume} {14}},\ \bibinfo {pages}
  {447} (\bibinfo {year} {2018})}\BibitemShut {NoStop}%
\bibitem [{\citenamefont {Carleo}\ \emph {et~al.}(2018)\citenamefont {Carleo},
  \citenamefont {Nomura},\ and\ \citenamefont {Imada}}]{Carleo:2018tj}%
  \BibitemOpen
  \bibfield  {author} {\bibinfo {author} {\bibfnamefont {G.}~\bibnamefont
  {Carleo}}, \bibinfo {author} {\bibfnamefont {Y.}~\bibnamefont {Nomura}},\
  and\ \bibinfo {author} {\bibfnamefont {M.}~\bibnamefont {Imada}},\ }\bibfield
   {title} {\bibinfo {title} {Constructing exact representations of quantum
  many-body systems with deep neural networks},\ }\href
  {https://doi.org/10.1038/s41467-018-07520-3} {\bibfield  {journal} {\bibinfo
  {journal} {Nat. Commun.}\ }\textbf {\bibinfo {volume} {9}},\ \bibinfo {pages}
  {5322} (\bibinfo {year} {2018})}\BibitemShut {NoStop}%
\bibitem [{\citenamefont {Xin}\ \emph {et~al.}(2019)\citenamefont {Xin},
  \citenamefont {Lu}, \citenamefont {Cao}, \citenamefont {Anikeeva},
  \citenamefont {Lu}, \citenamefont {Li}, \citenamefont {Long},\ and\
  \citenamefont {Zeng}}]{Xin:2019vo}%
  \BibitemOpen
  \bibfield  {author} {\bibinfo {author} {\bibfnamefont {T.}~\bibnamefont
  {Xin}}, \bibinfo {author} {\bibfnamefont {S.}~\bibnamefont {Lu}}, \bibinfo
  {author} {\bibfnamefont {N.}~\bibnamefont {Cao}}, \bibinfo {author}
  {\bibfnamefont {G.}~\bibnamefont {Anikeeva}}, \bibinfo {author}
  {\bibfnamefont {D.}~\bibnamefont {Lu}}, \bibinfo {author} {\bibfnamefont
  {J.}~\bibnamefont {Li}}, \bibinfo {author} {\bibfnamefont {G.}~\bibnamefont
  {Long}},\ and\ \bibinfo {author} {\bibfnamefont {B.}~\bibnamefont {Zeng}},\
  }\bibfield  {title} {\bibinfo {title} {Local-measurement-based quantum state
  tomography via neural networks},\ }\href
  {https://doi.org/10.1038/s41534-019-0222-3} {\bibfield  {journal} {\bibinfo
  {journal} {npj Quantum Inf.}\ }\textbf {\bibinfo {volume} {5}},\ \bibinfo
  {pages} {109} (\bibinfo {year} {2019})}\BibitemShut {NoStop}%
\bibitem [{\citenamefont {Palmieri}\ \emph {et~al.}(2020)\citenamefont
  {Palmieri}, \citenamefont {Kovlakov}, \citenamefont {Bianchi}, \citenamefont
  {Yudin}, \citenamefont {Straupe}, \citenamefont {Biamonte},\ and\
  \citenamefont {Kulik}}]{Palmieri:2020vc}%
  \BibitemOpen
  \bibfield  {author} {\bibinfo {author} {\bibfnamefont {A.~M.}\ \bibnamefont
  {Palmieri}}, \bibinfo {author} {\bibfnamefont {E.}~\bibnamefont {Kovlakov}},
  \bibinfo {author} {\bibfnamefont {F.}~\bibnamefont {Bianchi}}, \bibinfo
  {author} {\bibfnamefont {D.}~\bibnamefont {Yudin}}, \bibinfo {author}
  {\bibfnamefont {S.}~\bibnamefont {Straupe}}, \bibinfo {author} {\bibfnamefont
  {J.~D.}\ \bibnamefont {Biamonte}},\ and\ \bibinfo {author} {\bibfnamefont
  {S.}~\bibnamefont {Kulik}},\ }\bibfield  {title} {\bibinfo {title}
  {Experimental neural network enhanced quantum tomography},\ }\href
  {https://doi.org/10.1038/s41534-020-0248-6} {\bibfield  {journal} {\bibinfo
  {journal} {npj Quantum Inf.}\ }\textbf {\bibinfo {volume} {6}},\ \bibinfo
  {pages} {20} (\bibinfo {year} {2020})}\BibitemShut {NoStop}%
\bibitem [{\citenamefont {Melkani}\ \emph {et~al.}(2020)\citenamefont
  {Melkani}, \citenamefont {Gneiting},\ and\ \citenamefont
  {Nori}}]{Melkani:2020wo}%
  \BibitemOpen
  \bibfield  {author} {\bibinfo {author} {\bibfnamefont {A.}~\bibnamefont
  {Melkani}}, \bibinfo {author} {\bibfnamefont {C.}~\bibnamefont {Gneiting}},\
  and\ \bibinfo {author} {\bibfnamefont {F.}~\bibnamefont {Nori}},\ }\bibfield
  {title} {\bibinfo {title} {Eigenstate extraction with neural-network
  tomography},\ }\href {https://doi.org/10.1103/PhysRevA.102.022412} {\bibfield
   {journal} {\bibinfo  {journal} {Phys. Rev. A}\ }\textbf {\bibinfo {volume}
  {102}},\ \bibinfo {pages} {022412} (\bibinfo {year} {2020})}\BibitemShut
  {NoStop}%
\bibitem [{\citenamefont {Lohani}\ \emph {et~al.}(2020)\citenamefont {Lohani},
  \citenamefont {Kirby}, \citenamefont {Brodsky}, \citenamefont {Danaci},\ and\
  \citenamefont {Glasser}}]{Lohani:2020vi}%
  \BibitemOpen
  \bibfield  {author} {\bibinfo {author} {\bibfnamefont {S.}~\bibnamefont
  {Lohani}}, \bibinfo {author} {\bibfnamefont {B.~T.}\ \bibnamefont {Kirby}},
  \bibinfo {author} {\bibfnamefont {M.}~\bibnamefont {Brodsky}}, \bibinfo
  {author} {\bibfnamefont {O.}~\bibnamefont {Danaci}},\ and\ \bibinfo {author}
  {\bibfnamefont {R.~T.}\ \bibnamefont {Glasser}},\ }\bibfield  {title}
  {\bibinfo {title} {Machine learning assisted quantum state estimation},\
  }\href {https://doi.org/10.1088/2632-2153/ab9a21} {\bibfield  {journal}
  {\bibinfo  {journal} {Mach. Learn.: Sci. Technol.}\ }\textbf {\bibinfo
  {volume} {1}},\ \bibinfo {pages} {035007} (\bibinfo {year}
  {2020})}\BibitemShut {NoStop}%
\bibitem [{\citenamefont {Liu}\ \emph {et~al.}(2020)\citenamefont {Liu},
  \citenamefont {Wang}, \citenamefont {Xue}, \citenamefont {Huang},
  \citenamefont {Fu}, \citenamefont {Qiang}, \citenamefont {Xu}, \citenamefont
  {Huang}, \citenamefont {Deng}, \citenamefont {Guo}, \citenamefont {Yang},\
  and\ \citenamefont {Wu}}]{Liu:2020tn}%
  \BibitemOpen
  \bibfield  {author} {\bibinfo {author} {\bibfnamefont {Y.}~\bibnamefont
  {Liu}}, \bibinfo {author} {\bibfnamefont {D.}~\bibnamefont {Wang}}, \bibinfo
  {author} {\bibfnamefont {S.}~\bibnamefont {Xue}}, \bibinfo {author}
  {\bibfnamefont {A.}~\bibnamefont {Huang}}, \bibinfo {author} {\bibfnamefont
  {X.}~\bibnamefont {Fu}}, \bibinfo {author} {\bibfnamefont {X.}~\bibnamefont
  {Qiang}}, \bibinfo {author} {\bibfnamefont {P.}~\bibnamefont {Xu}}, \bibinfo
  {author} {\bibfnamefont {H.-L.}\ \bibnamefont {Huang}}, \bibinfo {author}
  {\bibfnamefont {M.}~\bibnamefont {Deng}}, \bibinfo {author} {\bibfnamefont
  {C.}~\bibnamefont {Guo}}, \bibinfo {author} {\bibfnamefont {X.}~\bibnamefont
  {Yang}},\ and\ \bibinfo {author} {\bibfnamefont {J.}~\bibnamefont {Wu}},\
  }\bibfield  {title} {\bibinfo {title} {Variational quantum circuits for
  quantum state tomography},\ }\href
  {https://doi.org/10.1103/PhysRevA.101.052316} {\bibfield  {journal} {\bibinfo
   {journal} {Phys. Rev. A}\ }\textbf {\bibinfo {volume} {101}},\ \bibinfo
  {pages} {052316} (\bibinfo {year} {2020})}\BibitemShut {NoStop}%
\bibitem [{\citenamefont {Quek}\ \emph {et~al.}(2021)\citenamefont {Quek},
  \citenamefont {Fort},\ and\ \citenamefont {Ng}}]{Quek:2021uc}%
  \BibitemOpen
  \bibfield  {author} {\bibinfo {author} {\bibfnamefont {Y.}~\bibnamefont
  {Quek}}, \bibinfo {author} {\bibfnamefont {S.}~\bibnamefont {Fort}},\ and\
  \bibinfo {author} {\bibfnamefont {H.~K.}\ \bibnamefont {Ng}},\ }\bibfield
  {title} {\bibinfo {title} {Adaptive quantum state tomography with neural
  networks},\ }\href {https://doi.org/10.1038/s41534-021-00436-9} {\bibfield
  {journal} {\bibinfo  {journal} {npj Quantum Inf.}\ }\textbf {\bibinfo
  {volume} {7}},\ \bibinfo {pages} {105} (\bibinfo {year} {2021})}\BibitemShut
  {NoStop}%
\bibitem [{\citenamefont {Carrasquilla}\ and\ \citenamefont
  {Torlai}(2021)}]{Carrasquilla:2021tx}%
  \BibitemOpen
  \bibfield  {author} {\bibinfo {author} {\bibfnamefont {J.}~\bibnamefont
  {Carrasquilla}}\ and\ \bibinfo {author} {\bibfnamefont {G.}~\bibnamefont
  {Torlai}},\ }\bibfield  {title} {\bibinfo {title} {How to use neural networks
  to investigate quantum many-body physics},\ }\href
  {https://doi.org/10.1103/PRXQuantum.2.040201} {\bibfield  {journal} {\bibinfo
   {journal} {PRX Quantum}\ }\textbf {\bibinfo {volume} {2}},\ \bibinfo {pages}
  {040201} (\bibinfo {year} {2021})}\BibitemShut {NoStop}%
\bibitem [{\citenamefont {Carrasquilla}\ \emph {et~al.}(2019)\citenamefont
  {Carrasquilla}, \citenamefont {Torlai}, \citenamefont {Melko},\ and\
  \citenamefont {Aolita}}]{Carrasquilla:2019wm}%
  \BibitemOpen
  \bibfield  {author} {\bibinfo {author} {\bibfnamefont {J.}~\bibnamefont
  {Carrasquilla}}, \bibinfo {author} {\bibfnamefont {G.}~\bibnamefont
  {Torlai}}, \bibinfo {author} {\bibfnamefont {R.~G.}\ \bibnamefont {Melko}},\
  and\ \bibinfo {author} {\bibfnamefont {L.}~\bibnamefont {Aolita}},\
  }\bibfield  {title} {\bibinfo {title} {Reconstructing quantum states with
  generative models},\ }\href {https://doi.org/10.1038/s42256-019-0028-1}
  {\bibfield  {journal} {\bibinfo  {journal} {Nat. Mach. Intell.}\ }\textbf
  {\bibinfo {volume} {1}},\ \bibinfo {pages} {155} (\bibinfo {year}
  {2019})}\BibitemShut {NoStop}%
\bibitem [{\citenamefont {Lloyd}\ and\ \citenamefont
  {Weedbrook}(2018)}]{Lloyd:2018uw}%
  \BibitemOpen
  \bibfield  {author} {\bibinfo {author} {\bibfnamefont {S.}~\bibnamefont
  {Lloyd}}\ and\ \bibinfo {author} {\bibfnamefont {C.}~\bibnamefont
  {Weedbrook}},\ }\bibfield  {title} {\bibinfo {title} {Quantum generative
  adversarial learning},\ }\href
  {https://doi.org/10.1103/PhysRevLett.121.040502} {\bibfield  {journal}
  {\bibinfo  {journal} {Phys. Rev. Lett.}\ }\textbf {\bibinfo {volume} {121}},\
  \bibinfo {pages} {040502} (\bibinfo {year} {2018})}\BibitemShut {NoStop}%
\bibitem [{\citenamefont {Tiunov}\ \emph {et~al.}(2020)\citenamefont {Tiunov},
  \citenamefont {Tiunova~(Vyborova)}, \citenamefont {Ulanov}, \citenamefont
  {Lvovsky},\ and\ \citenamefont {Fedorov}}]{Tiunov:2020wm}%
  \BibitemOpen
  \bibfield  {author} {\bibinfo {author} {\bibfnamefont {E.~S.}\ \bibnamefont
  {Tiunov}}, \bibinfo {author} {\bibfnamefont {V.~V.}\ \bibnamefont
  {Tiunova~(Vyborova)}}, \bibinfo {author} {\bibfnamefont {A.~E.}\ \bibnamefont
  {Ulanov}}, \bibinfo {author} {\bibfnamefont {A.~I.}\ \bibnamefont
  {Lvovsky}},\ and\ \bibinfo {author} {\bibfnamefont {A.~K.}\ \bibnamefont
  {Fedorov}},\ }\bibfield  {title} {\bibinfo {title} {Experimental quantum
  homodyne tomography via machine learning},\ }\href
  {https://doi.org/10.1364/OPTICA.389482} {\bibfield  {journal} {\bibinfo
  {journal} {Optica}\ }\textbf {\bibinfo {volume} {7}},\ \bibinfo {pages} {448}
  (\bibinfo {year} {2020})}\BibitemShut {NoStop}%
\bibitem [{\citenamefont {Cai}\ and\ \citenamefont {Liu}(2018)}]{Cai:2018wn}%
  \BibitemOpen
  \bibfield  {author} {\bibinfo {author} {\bibfnamefont {Z.}~\bibnamefont
  {Cai}}\ and\ \bibinfo {author} {\bibfnamefont {J.}~\bibnamefont {Liu}},\
  }\bibfield  {title} {\bibinfo {title} {Approximating quantum many-body wave
  functions using artificial neural networks},\ }\href
  {https://doi.org/10.1103/PhysRevB.97.035116} {\bibfield  {journal} {\bibinfo
  {journal} {Phys. Rev. B}\ }\textbf {\bibinfo {volume} {97}},\ \bibinfo
  {pages} {035116} (\bibinfo {year} {2018})}\BibitemShut {NoStop}%
\bibitem [{\citenamefont {Cha}\ \emph {et~al.}(2021)\citenamefont {Cha},
  \citenamefont {Ginsparg}, \citenamefont {Wu}, \citenamefont {Carrasquilla},
  \citenamefont {McMahon},\ and\ \citenamefont {Kim}}]{Cha:2021wn}%
  \BibitemOpen
  \bibfield  {author} {\bibinfo {author} {\bibfnamefont {P.}~\bibnamefont
  {Cha}}, \bibinfo {author} {\bibfnamefont {P.}~\bibnamefont {Ginsparg}},
  \bibinfo {author} {\bibfnamefont {F.}~\bibnamefont {Wu}}, \bibinfo {author}
  {\bibfnamefont {J.}~\bibnamefont {Carrasquilla}}, \bibinfo {author}
  {\bibfnamefont {P.~L.}\ \bibnamefont {McMahon}},\ and\ \bibinfo {author}
  {\bibfnamefont {E.-A.}\ \bibnamefont {Kim}},\ }\bibfield  {title} {\bibinfo
  {title} {Attention-based quantum tomography},\ }\href
  {https://doi.org/10.1088/2632-2153/ac362b} {\bibfield  {journal} {\bibinfo
  {journal} {Mach. Learn.: Sci. Technol.}\ }\textbf {\bibinfo {volume} {3}},\
  \bibinfo {pages} {01LT01} (\bibinfo {year} {2021})}\BibitemShut {NoStop}%
\bibitem [{\citenamefont {Kingma}\ and\ \citenamefont
  {Welling}(2019)}]{Kingma:2019uw}%
  \BibitemOpen
  \bibfield  {author} {\bibinfo {author} {\bibfnamefont {D.}~\bibnamefont
  {Kingma}}\ and\ \bibinfo {author} {\bibfnamefont {M.}~\bibnamefont
  {Welling}},\ }\bibfield  {title} {\bibinfo {title} {An introduction to
  variational autoencoders},\ }\href {https://doi.org/10.1561/2200000056}
  {\bibfield  {journal} {\bibinfo  {journal} {Found. Trends Mach. Learn.}\
  }\textbf {\bibinfo {volume} {12}},\ \bibinfo {pages} {307} (\bibinfo {year}
  {2019})}\BibitemShut {NoStop}%
\bibitem [{\citenamefont {Goodfellow}\ \emph {et~al.}(2014)\citenamefont
  {Goodfellow}, \citenamefont {Pouget-Abadie}, \citenamefont {Mirza},
  \citenamefont {Xu}, \citenamefont {Warde-Farley}, \citenamefont {Ozair},
  \citenamefont {Courville},\ and\ \citenamefont {Bengio}}]{Goodfellow:2014td}%
  \BibitemOpen
  \bibfield  {author} {\bibinfo {author} {\bibfnamefont {I.~J.}\ \bibnamefont
  {Goodfellow}}, \bibinfo {author} {\bibfnamefont {J.}~\bibnamefont
  {Pouget-Abadie}}, \bibinfo {author} {\bibfnamefont {M.}~\bibnamefont
  {Mirza}}, \bibinfo {author} {\bibfnamefont {B.}~\bibnamefont {Xu}}, \bibinfo
  {author} {\bibfnamefont {D.}~\bibnamefont {Warde-Farley}}, \bibinfo {author}
  {\bibfnamefont {S.}~\bibnamefont {Ozair}}, \bibinfo {author} {\bibfnamefont
  {A.}~\bibnamefont {Courville}},\ and\ \bibinfo {author} {\bibfnamefont
  {Y.}~\bibnamefont {Bengio}},\ }\bibfield  {title} {\bibinfo {title}
  {Generative adversarial nets},\ }in\ \href@noop {} {\emph {\bibinfo
  {booktitle} {Advances in Neural Information Processing Systems}}},\
  Vol.~\bibinfo {volume} {3}\ (\bibinfo  {publisher} {Curran Associates},\
  \bibinfo {address} {Montreal},\ \bibinfo {year} {2014})\ pp.\ \bibinfo
  {pages} {2672--2680}\BibitemShut {NoStop}%
\bibitem [{\citenamefont {Rocchetto}\ \emph {et~al.}(2018)\citenamefont
  {Rocchetto}, \citenamefont {Grant}, \citenamefont {Strelchuk}, \citenamefont
  {Carleo},\ and\ \citenamefont {Severini}}]{Rocchetto:2018wg}%
  \BibitemOpen
  \bibfield  {author} {\bibinfo {author} {\bibfnamefont {A.}~\bibnamefont
  {Rocchetto}}, \bibinfo {author} {\bibfnamefont {E.}~\bibnamefont {Grant}},
  \bibinfo {author} {\bibfnamefont {S.}~\bibnamefont {Strelchuk}}, \bibinfo
  {author} {\bibfnamefont {G.}~\bibnamefont {Carleo}},\ and\ \bibinfo {author}
  {\bibfnamefont {S.}~\bibnamefont {Severini}},\ }\bibfield  {title} {\bibinfo
  {title} {Learning hard quantum distributions with variational autoencoders},\
  }\href {https://doi.org/10.1038/s41534-018-0077-z} {\bibfield  {journal}
  {\bibinfo  {journal} {npj Quantum Inf.}\ }\textbf {\bibinfo {volume} {4}},\
  \bibinfo {pages} {28} (\bibinfo {year} {2018})}\BibitemShut {NoStop}%
\bibitem [{\citenamefont {Zoufal}\ \emph {et~al.}(2019)\citenamefont {Zoufal},
  \citenamefont {Lucchi},\ and\ \citenamefont {Woerner}}]{Zoufal:2019wi}%
  \BibitemOpen
  \bibfield  {author} {\bibinfo {author} {\bibfnamefont {C.}~\bibnamefont
  {Zoufal}}, \bibinfo {author} {\bibfnamefont {A.}~\bibnamefont {Lucchi}},\
  and\ \bibinfo {author} {\bibfnamefont {S.}~\bibnamefont {Woerner}},\
  }\bibfield  {title} {\bibinfo {title} {Quantum generative adversarial
  networks for learning and loading random distributions},\ }\href
  {https://doi.org/10.1038/s41534-019-0223-2} {\bibfield  {journal} {\bibinfo
  {journal} {npj Quantum Inf.}\ }\textbf {\bibinfo {volume} {5}},\ \bibinfo
  {pages} {103} (\bibinfo {year} {2019})}\BibitemShut {NoStop}%
\bibitem [{\citenamefont {Ahmed}\ \emph
  {et~al.}(2021{\natexlab{a}})\citenamefont {Ahmed}, \citenamefont
  {S{\'a}nchez~Mu{\~n}oz}, \citenamefont {Nori},\ and\ \citenamefont
  {Kockum}}]{Ahmed:2021vs}%
  \BibitemOpen
  \bibfield  {author} {\bibinfo {author} {\bibfnamefont {S.}~\bibnamefont
  {Ahmed}}, \bibinfo {author} {\bibfnamefont {C.}~\bibnamefont
  {S{\'a}nchez~Mu{\~n}oz}}, \bibinfo {author} {\bibfnamefont {F.}~\bibnamefont
  {Nori}},\ and\ \bibinfo {author} {\bibfnamefont {A.~F.}\ \bibnamefont
  {Kockum}},\ }\bibfield  {title} {\bibinfo {title} {Quantum state tomography
  with conditional generative adversarial networks},\ }\href
  {https://doi.org/10.1103/PhysRevLett.127.140502} {\bibfield  {journal}
  {\bibinfo  {journal} {Phys. Rev. Lett.}\ }\textbf {\bibinfo {volume} {127}},\
  \bibinfo {pages} {140502} (\bibinfo {year} {2021}{\natexlab{a}})}\BibitemShut
  {NoStop}%
\bibitem [{\citenamefont {Ahmed}\ \emph
  {et~al.}(2021{\natexlab{b}})\citenamefont {Ahmed}, \citenamefont
  {S{\'a}nchez~Mu{\~n}oz}, \citenamefont {Nori},\ and\ \citenamefont
  {Kockum}}]{Ahmed:2021vy}%
  \BibitemOpen
  \bibfield  {author} {\bibinfo {author} {\bibfnamefont {S.}~\bibnamefont
  {Ahmed}}, \bibinfo {author} {\bibfnamefont {C.}~\bibnamefont
  {S{\'a}nchez~Mu{\~n}oz}}, \bibinfo {author} {\bibfnamefont {F.}~\bibnamefont
  {Nori}},\ and\ \bibinfo {author} {\bibfnamefont {A.~F.}\ \bibnamefont
  {Kockum}},\ }\bibfield  {title} {\bibinfo {title} {Classification and
  reconstruction of optical quantum states with deep neural networks},\ }\href
  {https://doi.org/10.1103/PhysRevResearch.3.033278} {\bibfield  {journal}
  {\bibinfo  {journal} {Phys. Rev. Research}\ }\textbf {\bibinfo {volume}
  {3}},\ \bibinfo {pages} {033278} (\bibinfo {year}
  {2021}{\natexlab{b}})}\BibitemShut {NoStop}%
\bibitem [{\citenamefont {Helstrom}(1976)}]{Helstrom:1976ij}%
  \BibitemOpen
  \bibfield  {author} {\bibinfo {author} {\bibfnamefont {C.~W.}\ \bibnamefont
  {Helstrom}},\ }\href@noop {} {\emph {\bibinfo {title} {Quantum Detection and
  Estimation Theory}}}\ (\bibinfo  {publisher} {Academic},\ \bibinfo {address}
  {New York},\ \bibinfo {year} {1976})\BibitemShut {NoStop}%
\bibitem [{\citenamefont {Kimura}(2003)}]{Kimura:2003wz}%
  \BibitemOpen
  \bibfield  {author} {\bibinfo {author} {\bibfnamefont {G.}~\bibnamefont
  {Kimura}},\ }\bibfield  {title} {\bibinfo {title} {The {Bloch} vector for
  $n$-level systems},\ }\href
  {https://doi.org/https://doi.org/10.1016/S0375-9601(03)00941-1} {\bibfield
  {journal} {\bibinfo  {journal} {Phys. Lett. A}\ }\textbf {\bibinfo {volume}
  {314}},\ \bibinfo {pages} {339} (\bibinfo {year} {2003})}\BibitemShut
  {NoStop}%
\bibitem [{\citenamefont {Bertlmann}\ and\ \citenamefont
  {Krammer}(2008)}]{Bertlmann:2008uh}%
  \BibitemOpen
  \bibfield  {author} {\bibinfo {author} {\bibfnamefont {R.~A.}\ \bibnamefont
  {Bertlmann}}\ and\ \bibinfo {author} {\bibfnamefont {P.}~\bibnamefont
  {Krammer}},\ }\bibfield  {title} {\bibinfo {title} {Bloch vectors for
  qudits},\ }\href {https://doi.org/10.1088/1751-8113/41/23/235303} {\bibfield
  {journal} {\bibinfo  {journal} {J. Phys. A: Math. Theo.}\ }\textbf {\bibinfo
  {volume} {41}},\ \bibinfo {pages} {235303} (\bibinfo {year}
  {2008})}\BibitemShut {NoStop}%
\bibitem [{\citenamefont {Menda{\v s}}(2008)}]{Mendas:2008uf}%
  \BibitemOpen
  \bibfield  {author} {\bibinfo {author} {\bibfnamefont {I.~P.}\ \bibnamefont
  {Menda{\v s}}},\ }\bibfield  {title} {\bibinfo {title} {Classification and
  time evolution of density matrices for a $n$-state system},\ }\href
  {https://aip.scitation.org/doi/abs/10.1063/1.2982276} {\bibfield  {journal}
  {\bibinfo  {journal} {J. Math. Phys.}\ }\textbf {\bibinfo {volume} {49}},\
  \bibinfo {pages} {092102} (\bibinfo {year} {2008})}\BibitemShut {NoStop}%
\bibitem [{\citenamefont {Br{\"u}ning}\ \emph {et~al.}(2012)\citenamefont
  {Br{\"u}ning}, \citenamefont {M{\"a}kel{\"a}}, \citenamefont {Messina},\ and\
  \citenamefont {Petruccione}}]{Bruning:2012vd}%
  \BibitemOpen
  \bibfield  {author} {\bibinfo {author} {\bibfnamefont {E.}~\bibnamefont
  {Br{\"u}ning}}, \bibinfo {author} {\bibfnamefont {H.}~\bibnamefont
  {M{\"a}kel{\"a}}}, \bibinfo {author} {\bibfnamefont {A.}~\bibnamefont
  {Messina}},\ and\ \bibinfo {author} {\bibfnamefont {F.}~\bibnamefont
  {Petruccione}},\ }\bibfield  {title} {\bibinfo {title} {Parametrizations of
  density matrices},\ }\href {https://doi.org/10.1080/09500340.2011.632097}
  {\bibfield  {journal} {\bibinfo  {journal} {J. Mod. Opt.}\ }\textbf {\bibinfo
  {volume} {59}},\ \bibinfo {pages} {1} (\bibinfo {year} {2012})}\BibitemShut
  {NoStop}%
\bibitem [{\citenamefont {Penrose}(1955)}]{Penrose:1955aa}%
  \BibitemOpen
  \bibfield  {author} {\bibinfo {author} {\bibfnamefont {R.}~\bibnamefont
  {Penrose}},\ }\bibfield  {title} {\bibinfo {title} {A generalized inverse for
  matrices},\ }\href {https://doi.org/DOI: 10.1017/S0305004100030401}
  {\bibfield  {journal} {\bibinfo  {journal} {Proc. Cambridge Phil. Soc.}\
  }\textbf {\bibinfo {volume} {51}},\ \bibinfo {pages} {406} (\bibinfo {year}
  {1955})}\BibitemShut {NoStop}%
\bibitem [{\citenamefont {Ben-Israel}\ and\ \citenamefont
  {Greville}(1977)}]{Ben-Israel:1977aa}%
  \BibitemOpen
  \bibfield  {author} {\bibinfo {author} {\bibfnamefont {A.}~\bibnamefont
  {Ben-Israel}}\ and\ \bibinfo {author} {\bibfnamefont {T.~N.~E.}\ \bibnamefont
  {Greville}},\ }\href@noop {} {\emph {\bibinfo {title} {Generalized Inverses:
  Theory and Applications}}}\ (\bibinfo  {publisher} {Wiley},\ \bibinfo
  {address} {New York},\ \bibinfo {year} {1977})\BibitemShut {NoStop}%
\bibitem [{\citenamefont {Campbell}\ and\ \citenamefont
  {Meyer}(1991)}]{Campbell:1991aa}%
  \BibitemOpen
  \bibfield  {author} {\bibinfo {author} {\bibfnamefont {S.~L.}\ \bibnamefont
  {Campbell}}\ and\ \bibinfo {author} {\bibfnamefont {C.~D.~J.}\ \bibnamefont
  {Meyer}},\ }\href@noop {} {\emph {\bibinfo {title} {Generalized Inverses of
  Linear Transformations}}}\ (\bibinfo  {publisher} {Dover},\ \bibinfo
  {address} {New York},\ \bibinfo {year} {1991})\BibitemShut {NoStop}%
\bibitem [{\citenamefont {Lawson}\ and\ \citenamefont
  {Hanson}(1974)}]{Lawson:1974aa}%
  \BibitemOpen
  \bibfield  {author} {\bibinfo {author} {\bibfnamefont {C.}~\bibnamefont
  {Lawson}}\ and\ \bibinfo {author} {\bibfnamefont {R.}~\bibnamefont
  {Hanson}},\ }\href@noop {} {\emph {\bibinfo {title} {Solving Least Squares
  Problems}}}\ (\bibinfo  {publisher} {Prentice-Hall},\ \bibinfo {address}
  {Englewood Cliffs},\ \bibinfo {year} {1974})\BibitemShut {NoStop}%
\bibitem [{\citenamefont {Kaznady}\ and\ \citenamefont
  {James}(2009)}]{Kaznady:2009wf}%
  \BibitemOpen
  \bibfield  {author} {\bibinfo {author} {\bibfnamefont {M.~S.}\ \bibnamefont
  {Kaznady}}\ and\ \bibinfo {author} {\bibfnamefont {D.~F.~V.}\ \bibnamefont
  {James}},\ }\bibfield  {title} {\bibinfo {title} {Numerical strategies for
  quantum tomography: Alternatives to full optimization},\ }\href
  {https://doi.org/10.1103/PhysRevA.79.022109} {\bibfield  {journal} {\bibinfo
  {journal} {Physical Review A}\ }\textbf {\bibinfo {volume} {79}},\ \bibinfo
  {pages} {022109} (\bibinfo {year} {2009})}\BibitemShut {NoStop}%
\bibitem [{\citenamefont {Opatrn{\'y}}\ \emph {et~al.}(1997)\citenamefont
  {Opatrn{\'y}}, \citenamefont {Welsch},\ and\ \citenamefont
  {Vogel}}]{Opatrny:1997vv}%
  \BibitemOpen
  \bibfield  {author} {\bibinfo {author} {\bibfnamefont {T.}~\bibnamefont
  {Opatrn{\'y}}}, \bibinfo {author} {\bibfnamefont {D.~G.}\ \bibnamefont
  {Welsch}},\ and\ \bibinfo {author} {\bibfnamefont {W.}~\bibnamefont
  {Vogel}},\ }\bibfield  {title} {\bibinfo {title} {Least-squares inversion for
  density-matrix reconstruction},\ }\href
  {https://doi.org/10.1103/PhysRevA.56.1788} {\bibfield  {journal} {\bibinfo
  {journal} {Phys. Rev. A}\ }\textbf {\bibinfo {volume} {56}},\ \bibinfo
  {pages} {1788} (\bibinfo {year} {1997})}\BibitemShut {NoStop}%
\bibitem [{\citenamefont {Hallin}(2006)}]{Hallin:2006aa}%
  \BibitemOpen
  \bibfield  {author} {\bibinfo {author} {\bibfnamefont {M.}~\bibnamefont
  {Hallin}},\ }\bibinfo {title} {Gauss--Markov theorem in statistics},\ in\
  \href {https://doi.org/10.1002/9781118445112.stat07536} {\emph {\bibinfo
  {booktitle} {Encyclopedia of Environmetrics}}}\ (\bibinfo  {publisher} {John
  Wiley},\ \bibinfo {year} {2006})\BibitemShut {NoStop}%
\bibitem [{\citenamefont {Kay}(1993)}]{Kay:1993aa}%
  \BibitemOpen
  \bibfield  {author} {\bibinfo {author} {\bibfnamefont {S.~M.}\ \bibnamefont
  {Kay}},\ }\href@noop {} {\emph {\bibinfo {title} {Fundamentals of Statistical
  Signal Processing}}},\ Vol.~\bibinfo {volume} {1}\ (\bibinfo  {publisher}
  {Prentice Hall},\ \bibinfo {address} {Upper Saddle River},\ \bibinfo {year}
  {1993})\BibitemShut {NoStop}%
\bibitem [{\citenamefont {{\v R}eh{\'a}{\v c}ek}\ \emph
  {et~al.}(2007)\citenamefont {{\v R}eh{\'a}{\v c}ek}, \citenamefont {Hradil},
  \citenamefont {Knill},\ and\ \citenamefont {Lvovsky}}]{Rehacek:2007wr}%
  \BibitemOpen
  \bibfield  {author} {\bibinfo {author} {\bibfnamefont {J.}~\bibnamefont {{\v
  R}eh{\'a}{\v c}ek}}, \bibinfo {author} {\bibfnamefont {Z.}~\bibnamefont
  {Hradil}}, \bibinfo {author} {\bibfnamefont {E.}~\bibnamefont {Knill}},\ and\
  \bibinfo {author} {\bibfnamefont {A.~I.}\ \bibnamefont {Lvovsky}},\
  }\bibfield  {title} {\bibinfo {title} {Diluted maximum-likelihood algorithm
  for quantum tomography},\ }\href {https://doi.org/10.1103/PhysRevA.75.042108}
  {\bibfield  {journal} {\bibinfo  {journal} {Phys. Rev. A}\ }\textbf {\bibinfo
  {volume} {75}},\ \bibinfo {pages} {042108} (\bibinfo {year}
  {2007})}\BibitemShut {NoStop}%
\bibitem [{\citenamefont {Watkins}(1991)}]{Watkins:1991tc}%
  \BibitemOpen
  \bibfield  {author} {\bibinfo {author} {\bibfnamefont {D.}~\bibnamefont
  {Watkins}},\ }\href@noop {} {\emph {\bibinfo {title} {Fundamentals of Matrix
  Computations}}}\ (\bibinfo  {publisher} {Wiley},\ \bibinfo {address} {New
  York},\ \bibinfo {year} {1991})\BibitemShut {NoStop}%
\bibitem [{\citenamefont {Rumelhart}\ \emph {et~al.}(1986)\citenamefont
  {Rumelhart}, \citenamefont {Hinton},\ and\ \citenamefont
  {Williams}}]{Rumelhart:1986we}%
  \BibitemOpen
  \bibfield  {author} {\bibinfo {author} {\bibfnamefont {D.~E.}\ \bibnamefont
  {Rumelhart}}, \bibinfo {author} {\bibfnamefont {G.~E.}\ \bibnamefont
  {Hinton}},\ and\ \bibinfo {author} {\bibfnamefont {R.~J.}\ \bibnamefont
  {Williams}},\ }\bibfield  {title} {\bibinfo {title} {Learning representations
  by back-propagating errors},\ }\href {https://doi.org/10.1038/323533a0}
  {\bibfield  {journal} {\bibinfo  {journal} {Nature}\ }\textbf {\bibinfo
  {volume} {323}},\ \bibinfo {pages} {533} (\bibinfo {year}
  {1986})}\BibitemShut {NoStop}%
\bibitem [{\citenamefont {Goodfellow}\ \emph {et~al.}(2016)\citenamefont
  {Goodfellow}, \citenamefont {Bengio},\ and\ \citenamefont
  {Courville}}]{Goodfellow:2016wc}%
  \BibitemOpen
  \bibfield  {author} {\bibinfo {author} {\bibfnamefont {I.}~\bibnamefont
  {Goodfellow}}, \bibinfo {author} {\bibfnamefont {Y.}~\bibnamefont {Bengio}},\
  and\ \bibinfo {author} {\bibfnamefont {A.}~\bibnamefont {Courville}},\
  }\href@noop {} {\emph {\bibinfo {title} {Deep Learning}}}\ (\bibinfo
  {publisher} {MIT Press},\ \bibinfo {address} {Cambridge, MA},\ \bibinfo
  {year} {2016})\BibitemShut {NoStop}%
\bibitem [{\citenamefont {Ruder}(2016)}]{Ruder:2016tr}%
  \BibitemOpen
  \bibfield  {author} {\bibinfo {author} {\bibfnamefont {S.}~\bibnamefont
  {Ruder}},\ }\href@noop {} {\bibinfo {title} {An overview of gradient descent
  optimization algorithms}},\ \bibinfo {howpublished} {arXiv:1609.04747}
  (\bibinfo {year} {2016})\BibitemShut {NoStop}%
\bibitem [{\citenamefont {Kingma}\ and\ \citenamefont
  {Ba}(2014)}]{Kingma:2014us}%
  \BibitemOpen
  \bibfield  {author} {\bibinfo {author} {\bibfnamefont {D.~P.}\ \bibnamefont
  {Kingma}}\ and\ \bibinfo {author} {\bibfnamefont {J.}~\bibnamefont {Ba}},\
  }\href {https://doi.org/10.48550/ARXIV.1412.6980} {\bibinfo {title} {Adam: A
  method for stochastic optimization}} (\bibinfo {year} {2014})\BibitemShut
  {NoStop}%
\bibitem [{\citenamefont {Dozat}(2015)}]{Dozat:2015wi}%
  \BibitemOpen
  \bibfield  {author} {\bibinfo {author} {\bibfnamefont {T.}~\bibnamefont
  {Dozat}},\ }\href {T. Dozat, Incorporating Nesterov momentum into Adam,
  http://cs229.stanford.edu/proj2015/054_report.pdf} {\bibinfo {title}
  {Incorporating Nesterov momentum into Adam,}} (\bibinfo {year}
  {2015})\BibitemShut {NoStop}%
\bibitem [{\citenamefont {Dogo}\ \emph {et~al.}(2018)\citenamefont {Dogo},
  \citenamefont {Afolabi}, \citenamefont {Nwulu}, \citenamefont {Twala},\ and\
  \citenamefont {Aigbavboa}}]{Dogo:2018vp}%
  \BibitemOpen
  \bibfield  {author} {\bibinfo {author} {\bibfnamefont {E.~M.}\ \bibnamefont
  {Dogo}}, \bibinfo {author} {\bibfnamefont {O.~J.}\ \bibnamefont {Afolabi}},
  \bibinfo {author} {\bibfnamefont {N.~I.}\ \bibnamefont {Nwulu}}, \bibinfo
  {author} {\bibfnamefont {B.}~\bibnamefont {Twala}},\ and\ \bibinfo {author}
  {\bibfnamefont {C.~O.}\ \bibnamefont {Aigbavboa}},\ }\bibfield  {title}
  {\bibinfo {title} {A comparative analysis of gradient descent-based
  optimization algorithms on convolutional neural networks},\ }in\ \href
  {https://doi.org/10.1109/CTEMS.2018.8769211} {\emph {\bibinfo {booktitle}
  {2018 International Conference on Computational Techniques, Electronics and
  Mechanical Systems (CTEMS)}}}\ (\bibinfo {year} {2018})\ pp.\ \bibinfo
  {pages} {92--99}\BibitemShut {NoStop}%
\bibitem [{\citenamefont {Chollet}(2015)}]{Chollet:2015wy}%
  \BibitemOpen
  \bibfield  {author} {\bibinfo {author} {\bibfnamefont {F.}~\bibnamefont
  {Chollet}},\ }\href {https://github.com/fchollet/keras} {\bibinfo {title}
  {Keras}} (\bibinfo {year} {2015})\BibitemShut {NoStop}%
\bibitem [{\citenamefont {\emph{et al}}(2015)}]{Abadi:2015te}%
  \BibitemOpen
  \bibfield  {author} {\bibinfo {author} {\bibfnamefont {M.~A.}\ \bibnamefont
  {\emph{et al}}},\ }\href {http://tensorflow.org/} {\bibinfo {title}
  {{TensorFlow}: Large-scale machine learning on heterogeneous systems}}
  (\bibinfo {year} {2015}),\ \bibinfo {note} {software available from
  tensorflow.org}\BibitemShut {NoStop}%
\bibitem [{Kou()}]{Koutny:2022qp}%
  \BibitemOpen
  \href@noop {} {}\bibinfo {howpublished}
  {\url{https://github.com/dkoutny/ReconstructionOfQuantumStatesViaNNs}}\BibitemShut
  {NoStop}%
\bibitem [{\citenamefont {Osipov}\ \emph {et~al.}(2010)\citenamefont {Osipov},
  \citenamefont {Sommers},\ and\ \citenamefont
  {{\.Z}yczkowski}}]{Osipov:2010ux}%
  \BibitemOpen
  \bibfield  {author} {\bibinfo {author} {\bibfnamefont {V.~.~A.}\ \bibnamefont
  {Osipov}}, \bibinfo {author} {\bibfnamefont {H.}~\bibnamefont {Sommers}},\
  and\ \bibinfo {author} {\bibfnamefont {K.}~\bibnamefont {{\.Z}yczkowski}},\
  }\bibfield  {title} {\bibinfo {title} {Random bures mixed states and the
  distribution of their purity},\ }\href
  {https://doi.org/10.1088/1751-8113/43/5/055302} {\bibfield  {journal}
  {\bibinfo  {journal} {J. Phys. A: Math. Theo.}\ }\textbf {\bibinfo {volume}
  {43}},\ \bibinfo {pages} {055302} (\bibinfo {year} {2010})}\BibitemShut
  {NoStop}%
\bibitem [{\citenamefont {Ginibre}(1965)}]{Ginibre:1965vc}%
  \BibitemOpen
  \bibfield  {author} {\bibinfo {author} {\bibfnamefont {J.}~\bibnamefont
  {Ginibre}},\ }\bibfield  {title} {\bibinfo {title} {Statistical ensembles of
  complex, quaternion, and real matrices},\ }\href
  {https://doi.org/10.1063/1.1704292} {\bibfield  {journal} {\bibinfo
  {journal} {J. Math. Phys.}\ }\textbf {\bibinfo {volume} {6}},\ \bibinfo
  {pages} {440} (\bibinfo {year} {1965})}\BibitemShut {NoStop}%
\bibitem [{\citenamefont {Johansson}\ \emph {et~al.}(2012)\citenamefont
  {Johansson}, \citenamefont {Nation},\ and\ \citenamefont
  {Nori}}]{Johansson:2012wf}%
  \BibitemOpen
  \bibfield  {author} {\bibinfo {author} {\bibfnamefont {J.~R.}\ \bibnamefont
  {Johansson}}, \bibinfo {author} {\bibfnamefont {P.~D.}\ \bibnamefont
  {Nation}},\ and\ \bibinfo {author} {\bibfnamefont {F.}~\bibnamefont {Nori}},\
  }\bibfield  {title} {\bibinfo {title} {{QuTiP}: An open-source {Python}
  framework for the dynamics of open quantum systems},\ }\href
  {https://doi.org/https://doi.org/10.1016/j.cpc.2012.02.021} {\bibfield
  {journal} {\bibinfo  {journal} {Comp. Phys. Commun.}\ }\textbf {\bibinfo
  {volume} {183}},\ \bibinfo {pages} {1760} (\bibinfo {year}
  {2012})}\BibitemShut {NoStop}%
\bibitem [{\citenamefont {{\.Z}yczkowski}\ \emph {et~al.}(2011)\citenamefont
  {{\.Z}yczkowski}, \citenamefont {Penson}, \citenamefont {Nechita},\ and\
  \citenamefont {Collins}}]{Zyczkowski:2011wo}%
  \BibitemOpen
  \bibfield  {author} {\bibinfo {author} {\bibfnamefont {K.}~\bibnamefont
  {{\.Z}yczkowski}}, \bibinfo {author} {\bibfnamefont {K.~A.}\ \bibnamefont
  {Penson}}, \bibinfo {author} {\bibfnamefont {I.}~\bibnamefont {Nechita}},\
  and\ \bibinfo {author} {\bibfnamefont {B.}~\bibnamefont {Collins}},\
  }\bibfield  {title} {\bibinfo {title} {Generating random density matrices},\
  }\href {https://doi.org/10.1063/1.3595693} {\bibfield  {journal} {\bibinfo
  {journal} {J. Math. Phys.}\ }\textbf {\bibinfo {volume} {52}},\ \bibinfo
  {pages} {062201} (\bibinfo {year} {2011})}\BibitemShut {NoStop}%
\bibitem [{\citenamefont {Hausladen}\ and\ \citenamefont
  {Wootters}(1994)}]{Hausladen:1994aa}%
  \BibitemOpen
  \bibfield  {author} {\bibinfo {author} {\bibfnamefont {P.}~\bibnamefont
  {Hausladen}}\ and\ \bibinfo {author} {\bibfnamefont {W.~K.}\ \bibnamefont
  {Wootters}},\ }\bibfield  {title} {\bibinfo {title} {A `pretty good'
  measurement for distinguishing quantum states},\ }\href
  {https://doi.org/10.1080/09500349414552221} {\bibfield  {journal} {\bibinfo
  {journal} {J. Mod. Opt.}\ }\textbf {\bibinfo {volume} {41}},\ \bibinfo
  {pages} {2385} (\bibinfo {year} {1994})}\BibitemShut {NoStop}%
\bibitem [{\citenamefont {Eldar}\ and\ \citenamefont
  {Forney}(2001)}]{Eldar:2001uw}%
  \BibitemOpen
  \bibfield  {author} {\bibinfo {author} {\bibfnamefont {Y.~C.}\ \bibnamefont
  {Eldar}}\ and\ \bibinfo {author} {\bibfnamefont {G.~D.}\ \bibnamefont
  {Forney}},\ }\bibfield  {title} {\bibinfo {title} {On quantum detection and
  the square-root measurement},\ }\href {https://doi.org/10.1109/18.915636}
  {\bibfield  {journal} {\bibinfo  {journal} {IEEE Trans. Inform. Theory}\
  }\textbf {\bibinfo {volume} {47}},\ \bibinfo {pages} {858} (\bibinfo {year}
  {2001})}\BibitemShut {NoStop}%
\bibitem [{\citenamefont {Dalla~Pozza}\ and\ \citenamefont
  {Pierobon}(2015)}]{Dalla-Pozza:2015aa}%
  \BibitemOpen
  \bibfield  {author} {\bibinfo {author} {\bibfnamefont {N.}~\bibnamefont
  {Dalla~Pozza}}\ and\ \bibinfo {author} {\bibfnamefont {G.}~\bibnamefont
  {Pierobon}},\ }\bibfield  {title} {\bibinfo {title} {Optimality of
  square-root measurements in quantum state discrimination},\ }\href
  {http://link.aps.org/doi/10.1103/PhysRevA.91.042334} {\bibfield  {journal}
  {\bibinfo  {journal} {Phys. Rev. A}\ }\textbf {\bibinfo {volume} {91}},\
  \bibinfo {pages} {042334} (\bibinfo {year} {2015})}\BibitemShut {NoStop}%
\end{thebibliography}
%

\end{document}